\RequirePackage{lineno}
\documentclass[ALICE,manyauthors]{cernphprep}
\usepackage[comma,square,numbers,sort&compress]{natbib}
\usepackage{graphicx}
\usepackage{epstopdf}
\usepackage{dcolumn}
\usepackage{bm}
\usepackage{color}
\usepackage{hyperref}
\usepackage{multirow}

\newcommand{\pbar}{$\rm\overline{p}$}
\newcommand{\dbar}{$\rm\overline{d}$}

\newcommand{\s}{$\sqrt{s}$}
\newcommand{\snn}{$\sqrt{s_{\mathrm{NN}}}$}
\newcommand{\pt}{$\ensuremath{p_{\rm T}}$}
\newcommand{\dedx}{d$E$/d$x$}
\newcommand{\dndy}{d$N$/d$y$}

\newcommand{\he}{$^{3}{\mathrm{He}}$}
\newcommand{\be}{$\begin{equation}$}
\newcommand{\ee}{$\end{equation}$}

\graphicspath{{figs/}}

\begin{document}

\begin{titlepage}
\PHyear{2015}
\PHnumber{025}      
\PHdate{02 Feb}  

\title{Production of light nuclei and anti-nuclei\\ in pp and Pb--Pb
  collisions at LHC energies}
\ShortTitle{Production of light (anti-)nuclei in pp and Pb--Pb
  collisions at LHC energies}
\Collaboration{ALICE Collaboration\thanks{See Appendix~\ref{app:collab} for the list of collaboration members}}
\ShortAuthor{ALICE Collaboration} 

\begin{abstract} 
The production of (anti-)deuteron and (anti-)\he\
nuclei in Pb--Pb collisions at \snn\ = 2.76~TeV has been studied
using the ALICE detector at the
LHC. The spectra exhibit a significant hardening with
increasing centrality. Combined blast-wave fits of several
particles support the interpretation that this behavior is caused
by an increase of radial flow.
The integrated particle yields are discussed
in the context of coalescence and thermal-statistical model
expectations. The particle ratios, 
\he/d 
and \he/p, in Pb--Pb
collisions are found to be in agreement with a common chemical
freeze-out temperature of $T_{\rm chem} \approx 156$ MeV. These
ratios do not vary with centrality which is in agreement with the
thermal-statistical model. In a coalescence approach, it excludes
models in which nucleus production is proportional to the
particle multiplicity and favors those in which it is proportional to the particle
density instead. In addition, the observation of 31 anti-tritons in Pb--Pb collisions is reported.
For comparison, the deuteron spectrum in pp collisions at \s\ = 7 TeV is also presented. 
While the p/$\pi$ ratio is similar in pp and Pb--Pb collisions, the d/p ratio in pp collisions is found to be 
lower by a factor of 2.2 than in Pb--Pb collisions.  

\end{abstract}
%
%
\end{titlepage}
\setcounter{page}{2}
%


\section{\label{secIntroduction}Introduction}

Collisions of ultra-relativistic ions create suitable conditions for
producing light (anti-)nuclei, because a high energy density is reached over a large volume.
Under these conditions, hot and dense matter, which contains approximately equal numbers of quarks
and anti-quarks at mid-rapidity, is produced for a short duration (a few $10^{-23}$ s).
The system cools down and undergoes a transition to
a hadron gas. 
While the hadronic yields are fixed at the moment when the rate of 
inelastic collisions becomes negligible (chemical freeze-out), 
the transverse momentum distributions continue to change
until also elastic interactions cease (kinetic freeze-out).

The formation of \mbox{(anti-)}nuclei is 
very sensitive to the chemical freeze-out conditions as well as to the dynamics
of the emitting source. 
The production scenarios are typically discussed within two approaches:
(i) The thermal-statistical approach has been very successful not
only in describing the integrated yield of the hadrons but also of composite
nuclei~\cite{Andronic:2010qu,Cleymans:2011pe,Andronic:2011yq}. In this picture,
the chemical freeze-out temperature $T_{\rm{chem}}$ (predicted around 160 MeV)
acts as the key parameter. The strong sensitivity of the abundance of nuclei to
the choice of $T_{\rm{chem}}$ is caused by their large mass $m$ and the
exponential dependence of the yield on the temperature given by $\exp(-m/T_{\rm chem})$.
(ii) In the coalescence model, nuclei are formed by protons and
neutrons which are nearby in phase space and exhibit similar
velocities~\cite{Butler:1963pp,Kapusta:1980zz}. A quantitative
description of this process is typically based on the coalescence
parameter $B_{A}$ 
and has
been applied to many collision systems at various energies
\cite{Afanasev:2000ku,Adler:2001uy,
Abelev:2009ae,Schael:2006fd,Aktas:2004pq,Abramov:1986ti,Anticic:2004yj,Bearden:2000we}.
The binding energy of light nuclei is very small (around few MeV), so they can hardly
remain intact during hadronic interactions, even if only quasi-elastic
scattering during the hadronic phase with temperatures between 100
MeV and 170 MeV is considered.
When produced thermally at chemical freeze-out, they might break apart and be created again by
final-state coalescence~\cite{Scheibl:1998tk}. It turns out that
both, the thermal approach and the coalescence mechanism, give very similar
predictions~\cite{Steinheimer:2012tb}.

The production of light nuclei has attracted attention already at
lower incident energies in heavy-ion collisions at the AGS, SPS,
and RHIC~\cite{Arsenescu:2003eg,Gutbrod:1988gt,Barrette:1994tw}.
A study of the dependence on \snn\ is of particular interest, because
different production mechanisms might dominate at various energies,
e.g.~a formation via spectator fragmentation at lower energies or
via coalescence/thermal mechanisms at higher ones.
In all cases, an exponential drop in the yield was found with
increasing mass of the
nuclei~\cite{BraunMunzinger:1994iq,Armstrong:2000gz}.
At RHIC and now at the LHC,
matter with a high content
of strange and of anti-quarks is created in heavy-ion
collisions. This has led to the first observation of
anti-alphas~\cite{Agakishiev:2011ib}
and of anti-hypertritons~\cite{Abelev:2010}. Their yields at LHC
energies were predicted based on thermal model estimates 
in~\cite{Andronic:2010qu,Cleymans:2011pe}.

In this paper, a detailed study of light (anti-)nuclei produced
in the mid-rapidity region in Pb--Pb collisions at \snn~= 2.76 TeV and a 
comparison with deuteron production in pp collisions at $\sqrt{s}$ = 7 TeV
using A Large Ion Collider Experiment (ALICE)~\cite{Aamodt:2008zz} is presented.
The paper is organized as follows:
In Section~\ref{secAnalysis}, details of the analysis technique used
to extract raw yields, acceptance and efficiency corrections of
\mbox{(anti-)}deuterons and \mbox{(anti-)}$\rm^3He$ are presented.
The results are given in Section~\ref{secResults} which starts
with a comparison of the production of nuclei and anti-nuclei
along with studies related to the hadronic interaction of 
anti-nuclei with the detector material.
Then, the transverse momentum spectra, \pt-integrated yields and
average transverse momenta are shown. The observation of
\mbox{(anti-)tritons} is also discussed in this section.
In Section~\ref{secDis}, the results
are discussed along
with a description using a blast-wave approach, and 
are compared with expectations from the thermal-statistical and
coalescence models. The measurement of \mbox{(anti-)}alphas and
\mbox{(anti-)}hypertritons will be shown in subsequent publications.

\section{\label{secAnalysis} Experiment and data analysis}

\subsection{The ALICE detector}
The results presented in this paper are obtained from the data
collected by 
the ALICE experiment at the LHC. Its performance and the
description of its various subsystems are discussed in detail in
Refs.~\cite{Aamodt:2008zz,Abelev:2014ffa}.
The ALICE detector has excellent particle
identification capabilities.
The main detectors used in this analysis are the Time Projection
Chamber (TPC)~\cite{Alme:2010ke}, the Time-Of-Flight detector
(TOF)~\cite{Akindinov:2010zzb}, and the Inner Tracking System
(ITS)~\cite{Aamodt:2010aa}. All detectors are positioned in a
solenoidal magnetic field of $B$ = 0.5 T. As the main tracking
device, the TPC provides full azimuthal acceptance for tracks in
the pseudo-rapidity region $|\eta| <$ 0.8. In addition, it
provides particle identification via the measurement of the
specific energy loss d$E$/d$x$.
It allows the identification of \mbox{(anti-)}\he\ over the entire momentum
range under study and the measurement is only limited by the
available statistics. The velocity information from the TOF
detector is in addition used to identify deuterons with transverse momenta (\pt)
above 1.4~GeV/$c$ and \mbox{(anti-)}tritons in the transverse momentum range of 
0.6~GeV/$c$ $<$ $p_{\mathrm{T}}$ $<$ 1.6 GeV/$c$.
The detector provides a similar acceptance as the TPC and its
total time resolution for tracks from Pb--Pb collisions corresponds to 
about 80 ps which is determined by the intrinsic time resolution of the
detector and the accuracy of the start time measurement.
By a combined analysis of TPC and TOF 
data, deuterons are identified up to 4.5~GeV/$c$ in Pb--Pb collisions.
In case of pp collisions, the less precisely determined start time leads to a time
resolution of about 120 ps and the identification is limited to about 3~GeV/$c$.
The precise space-point resolution in the six silicon layers of
the ITS allows a precise separation of primary and secondary
particles in the high track density
region close to the primary vertex.

\begin{figure}
 \begin{center}
  \includegraphics[width=9.5cm]{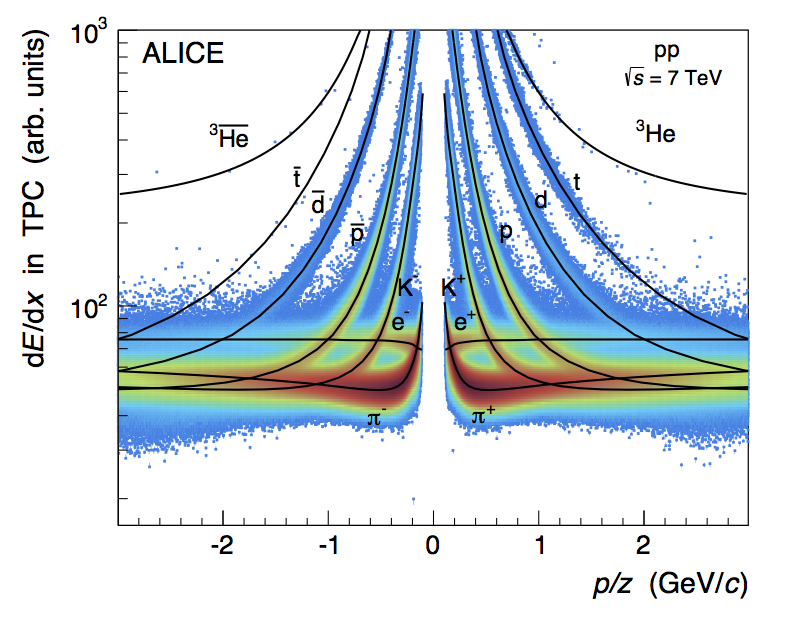}
   \includegraphics[width=9.5cm]{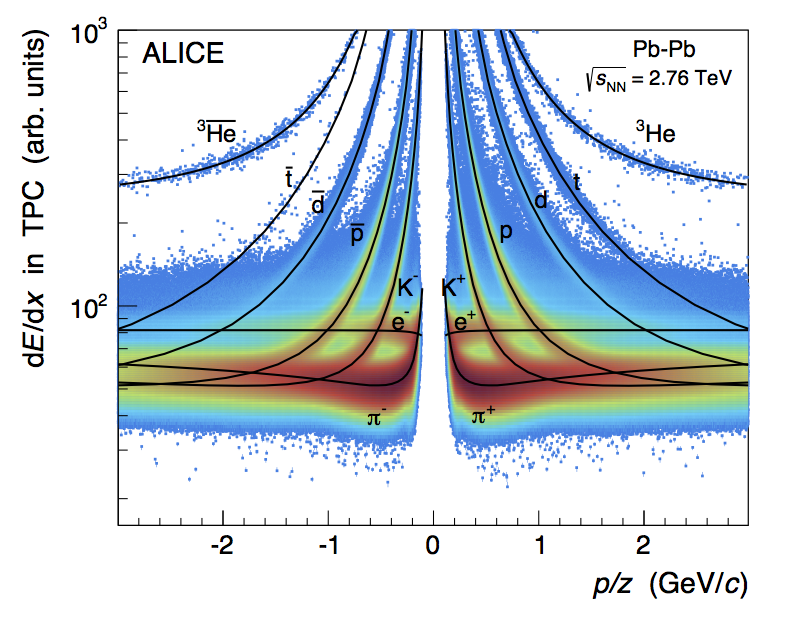}
  \caption{(Color online) Specific energy loss (d$E$/d$x$) vs.~rigidity
   (momentum/charge) for TPC tracks from pp collisions at \s\ = 7 TeV (top panel) and from 0-80\% most central Pb--Pb
   collisions at \snn\ = 2.76 TeV (bottom panel). The solid lines represent a parametrization of the Bethe-Bloch curve.}
  \label{dedx}
 \end{center}
\end{figure}

\begin{figure}
 \begin{center}
  \includegraphics[width=8.5cm]{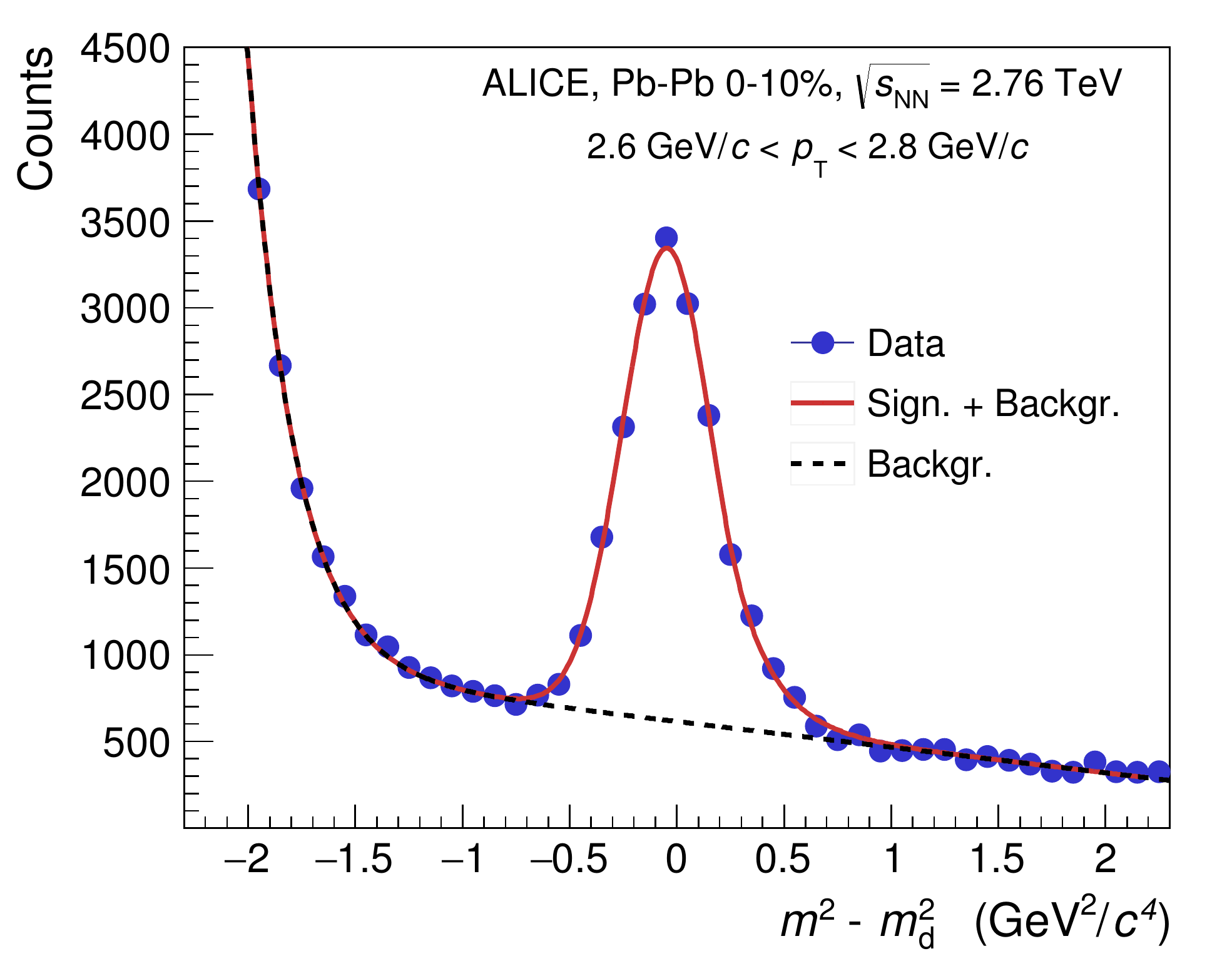}
  \caption{(Color online)
Distribution of $(m^{2} - m^{2}_{\rm d})$
  measured with the TOF detector for tracks with
  2.6 GeV/$c <~\pt\ < 2.8$ GeV/$c$ from central Pb--Pb collisions showing the
  peak corresponding to the deuteron mass $m_{\rm d}$ and the background from mismatched tracks (black dotted line) which is subtracted to obtain the raw yields (see text for details).}
  \label{tofPerformance}
 \end{center}
\end{figure}

\subsection{Event and track selection}

For this analysis, the data collected in the year 2010 are used.
In total, the data
sample consists of nearly $14$ million Pb--Pb collisions at \snn\ =
2.76~TeV and  $380$ million minimum-bias triggered events for pp
collisions at \s\ = 7~TeV after off-line event selection.

A pair of forward scintillator hodoscopes, the V0 detectors (2.8
$<$ $\eta$ $<$ 5.1 and -3.7 $<$ $\eta$ $<$ -1.7), measured the
arrival time of particles with a resolution of 1 ns and was used
for triggering purposes and for centrality determination of Pb--Pb
collisions. In pp collisions, the data were collected using a
minimum-bias trigger requiring at least one hit in either of the
V0 detectors or in the two innermost layers of the ITS (Silicon
Pixel Detector, SPD). The trigger condition during the Pb--Pb data
taking was changed with time to cope with the increasing
luminosity delivered by the LHC. It was restricted offline to a
homogenous condition, requiring at least two hits in the SPD and one hit in
either of the V0 detectors. This condition was shown to be fully
efficient for the 90\% most central events \cite{Abelev:2013qoq}.
A signal in the V0 was required to lie in a narrow time window
($\approx$ 30 ns) around the nominal collision time in order to
reject any contamination from beam-induced background. Only events
with a reconstructed primary vertex position in the fiducial
region $|V_{z}| < 10$~cm were considered in the analysis. The V0
amplitude distribution was also used to determine the centrality
of the heavy-ion collisions.
It was fitted with a
Glauber Monte-Carlo model to compute the fraction of the hadronic
cross section corresponding to a given range of V0 amplitude.
Based on those studies, the data were divided in several
centrality percentiles, selecting on signal amplitudes measured in
the V0 \cite{Abelev:2013qoq}.
The contamination from electromagnetic processes has been found to
be negligible
for the 80\% most central events. 

In this analysis, the production of primary deuterons and
$^{3}$He-nuclei as well as their respective anti-particles are
measured at mid-rapidity. In order to provide optimal particle
identification by reducing the difference between transverse and
total momentum, the spectra are provided within a rapidity window
of $|y| < 0.5$.
In addition, only those tracks in the full tracking acceptance of
$|\eta| < 0.8$ are selected.
The
extrapolation of the yield at low momenta, where the acceptance
does not cover the full $|y| < 0.5$ region, is done by assuming a
flat distribution in $y$ and by determining d$\eta$/d$y$ for each
\pt-interval. Primary particles are defined as prompt particles
produced in the collision including all decay products, except
products from weak decays of light flavor hadrons and of muons. In
order to select 
primary tracks of suitable quality, various track selection cuts
are applied. At least 70 clusters in the TPC and two points in the
ITS (out of which at least one in the SPD) are required. These selections
guarantee a track momentum resolution of 2\% in the relevant \pt-range
and a d$E$/d$x$ resolution of about~6\%, as well as a determination of
the Distance-of-Closest-Approach to the primary vertex in the
plane perpendicular~(DCA$_{xy}$) and parallel~(DCA$_{z}$) to the beam axis
with a resolution of better than 300~$\mu$m in the transverse direction~\cite{Abelev:2014ffa}. 
Furthermore, it is required that the  $\chi^2$ per TPC cluster is less than 4 and tracks of weak-decay products
are rejected as they cannot originate from the tracks of
primary nuclei.

\subsection{Particle identification}

Particle identification is mainly performed using the
TPC~\cite{Alme:2010ke}.
It is based on the measurement of the specific ionization energy
deposit
(\dedx) of charged particles.
Figure~\ref{dedx} shows the \dedx\ versus
rigidity (momentum/charge, $p/z$) of TPC tracks for pp collisions
at $\sqrt{s}$ = 7 TeV (top panel) and for Pb--Pb collisions at
$\sqrt{s_{\rm{NN}}}$ = 2.76 TeV (bottom panel).
Nuclei and anti-nuclei like (anti-)deuterons, (anti-)tritons, and
(anti-)$^{3}\rm{He}$ are clearly identified over a wide range of
momenta. The solid curves represent a parametrization of the
Bethe-Bloch function for the different particle species. In practice, it
is required that the measured energy-loss signal of a track
lies in a 3$\sigma$ window around
the expected value for a given mass
hypothesis. While this method provides a pure sample of $^{3}$He
nuclei in the  \pt-range between 2 GeV/$c$ and 7 GeV/$c$,
it is limited to about \pt~$<$~1.4~GeV/$c$ for deuterons.

In order to extend the \pt-reach of the deuteron measurement,
the TOF system is used above this momentum in addition. Based on the 
measured flight time $t$, the mass $m$ of a particle can be calculated as

\begin{equation}
	m^{2} = {p^{2} \over c^{2} } \cdot \Bigl({c^{2}t^{2} \over L^{2}} - 1 \Bigr)  \;,
\end{equation} 

\noindent where the total momentum $p$ and the track length $L$ are determined
with the tracking detectors. Figure~\ref{tofPerformance} shows the obtained 
$\Delta m^{2}$ distribution, where the deuteron mass
square ($m_{\rm d}^{2}$) was subtracted, for a \pt-bin between 2.6~GeV/$c$ 
and 2.8~GeV/$c$. For each \pt-bin, the $\Delta m^{2}$ distribution is fitted with a Gaussian function with
an exponential tail for the signal. Since the background mainly originates from two components,
namely wrong associations of a track with a TOF cluster and the non-Gaussian tail of lower mass particles,
it is described with a first order polynomial to which an exponential function is added.
The same procedure for signal extraction and background subtraction is applied in the analysis of pp collisions.

\subsection{Background rejection}

\begin{figure}
\begin{center}
 \includegraphics[width=7.7cm]{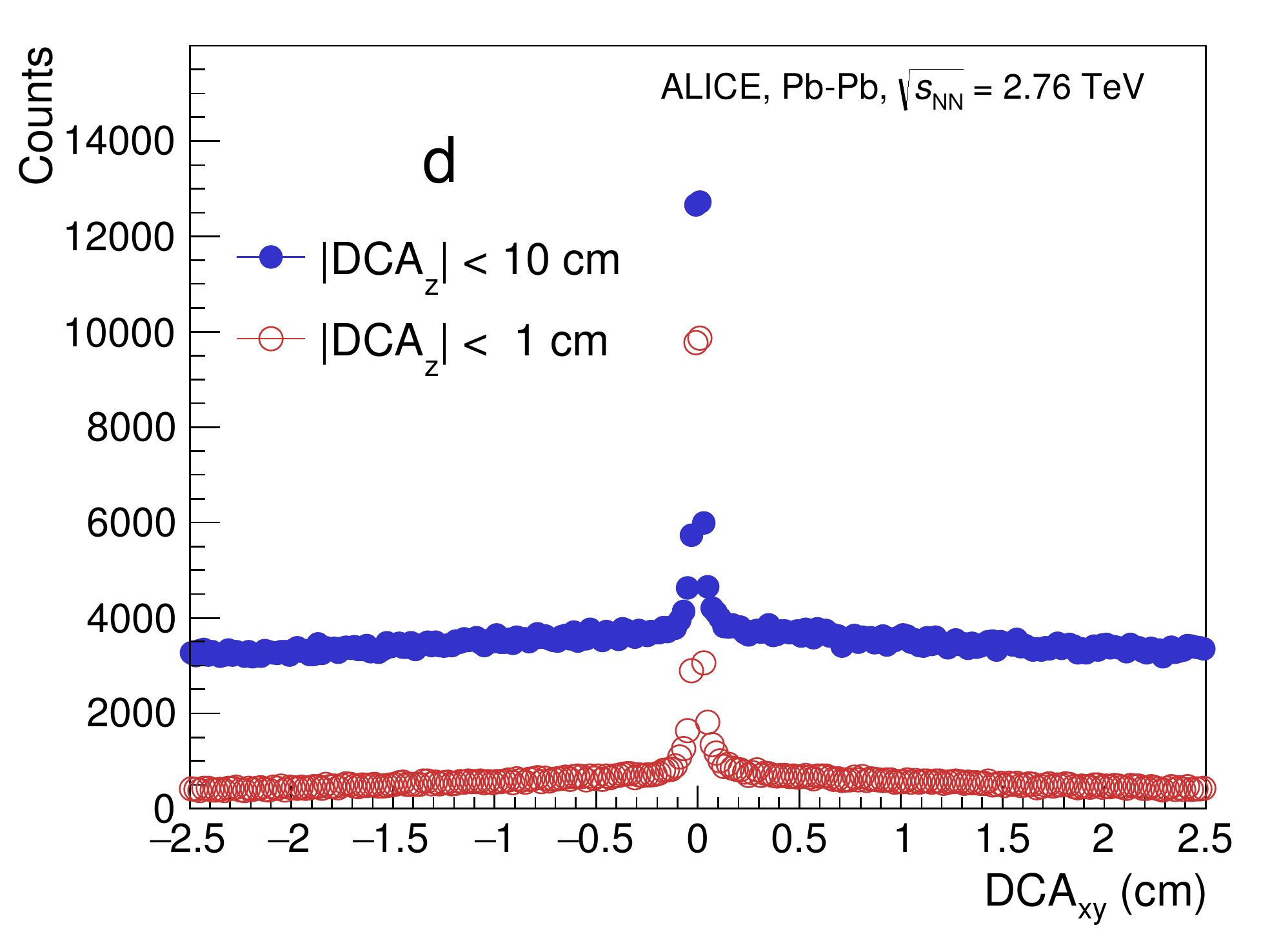}
\includegraphics[width=7.7cm]{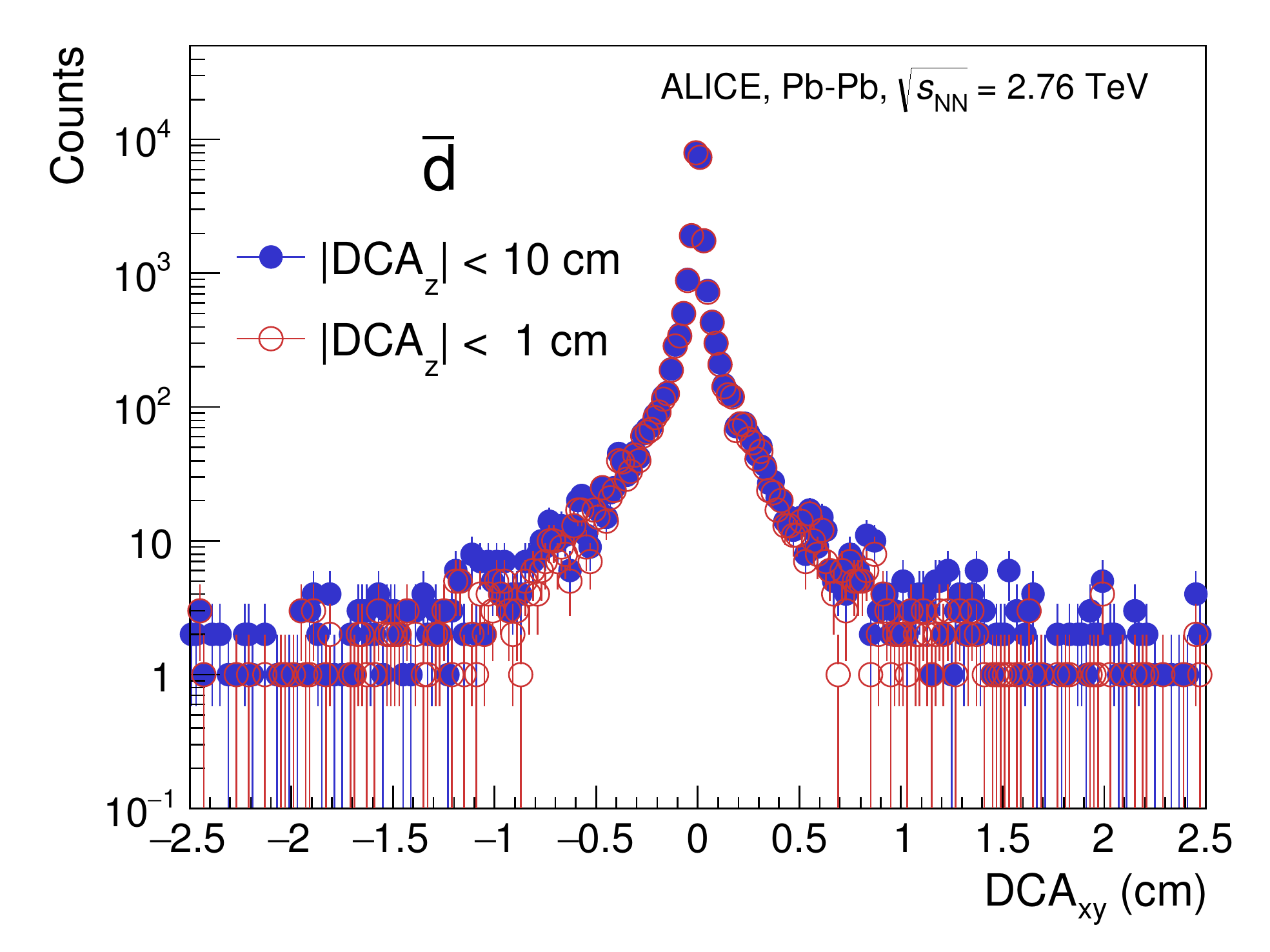}
\caption{(Color online) Distribution  of DCA$_{xy}$ for deuterons
 (left) and anti-deuterons (right) in the transverse momentum range
 $0.7$~GeV/$c$~$< \ensuremath{p_{\rm T}} < 1.4$~GeV/$c$ for 0-80\% most central Pb--Pb collisions 
 at \snn\ = 2.76 TeV demonstrating the influence of cuts in DCA$_{z}$
 on d and \dbar.}
\label{DCA_plot}
 \end{center}
\end{figure}

\begin{figure}
 \begin{center}
  \includegraphics[width=8.5cm]{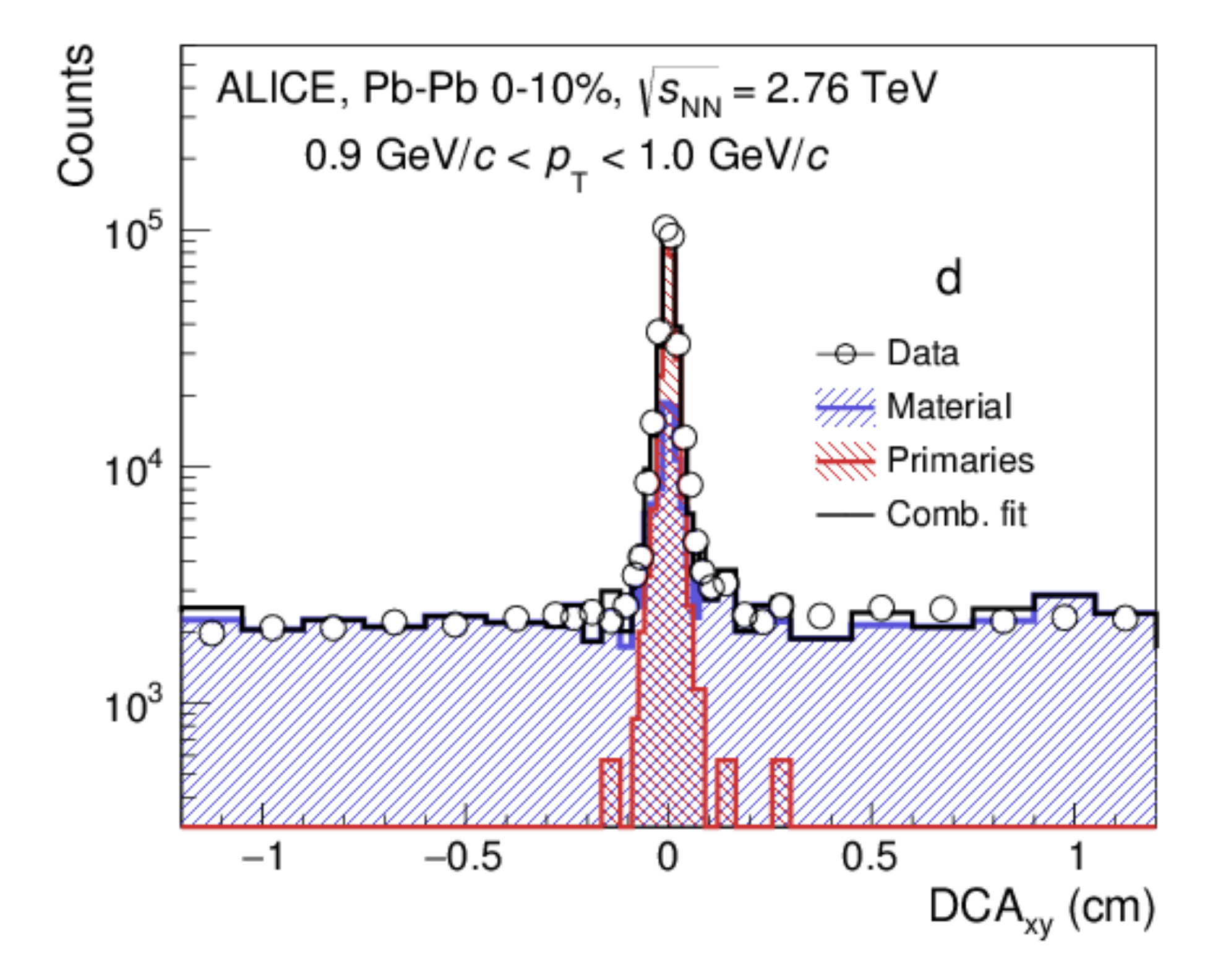}
 \caption{(Color online) Distribution of DCA$_{xy}$ of identified deuterons
 in the transverse momentum range $0.9$ GeV/$c$
 $< \ensuremath{p_{\rm T}} < 1.0$ GeV/$c$ for central Pb--Pb collisions
 (\snn\ = 2.76 TeV) along with the Monte-Carlo
 templates which are fitted to the data (see text for details).}
  \label{dcaxy}
\end{center}
\end{figure}

Particles produced in the collisions might interact with the
detector material and the beam pipe
which leads to the production of
secondary particles. The probability of anti-nucleus production
from the interaction of primary particles with detector material
is negligible, whereas the sample of nuclei may include primary as well as
secondary particles originating from the material. This contamination
is exponentially decreasing with increasing momentum. In addition,
it is about five times larger in central compared to peripheral Pb--Pb
or pp events because of the higher probability of a fake ITS hit
assignment to secondary tracks. Most of the secondary particles
from material have a large DCA to the primary vertex and hence
this information is used to correct
for the 
contamination. Figure~\ref{DCA_plot} shows the
DCA$_{xy}$ 
distribution for deuterons (left
panel) and anti-deuterons (right panel) for Pb--Pb collisions at
\snn\ = 2.76 TeV. The distributions are shown for two different
$|$DCA$_{z}|$ cuts. As can be seen from the figure, a strict
$|$DCA$_{z}|$ cut of 1.0 cm cuts a large fraction of background
for nuclei, but does not change the distribution for anti-nuclei.
At sufficiently high momenta (above 1.4 GeV/$c$ for deuterons and above 2
GeV/$c$ for \he),
the secondary and knock-out contamination caused by material
is in this way reduced to a negligible level 
and the raw yield can be directly extracted. In order to extend
the measurement of deuterons to lower momenta in Pb--Pb
collisions, the DCA$_{xy}$ distribution for deuterons in each
transverse momentum ($p_{\rm T}$)-interval was fitted with the
expected shapes (called ``templates'' in the following) as extracted
from Monte-Carlo events. Figure~\ref{dcaxy} shows a typical example of
this procedure for tracks with transverse momentum range $0.9$ GeV/$c$ $<
\ensuremath{p_{\rm T}} < 1.0$ GeV/$c$. One template for primary particles
and one template for secondary particles from material are used. The
characteristic shape of the template used for knock-out nuclei
from material with its flat behavior at large DCA$_{xy}$ allows a
precise distinction between the two contributions. The significant
peak at small $|$DCA$_{xy}|$ is caused by those knock-out nuclei
to which a cluster in one of the SPD layers is wrongly
associated. The obtained fraction of primary particles is then
used 
to calculate the raw yield in the corresponding $p_{\rm T}$-bin.
The same technique is applied for background rejection and raw yield
extraction of deuterons for pp collisions at \s~=~7~TeV.

\subsection{Efficiency and acceptance}
The final \pt-spectra of nuclei are obtained by correcting the raw
spectra for tracking efficiency and acceptance based on
Monte-Carlo (MC) generated events.
Standard event generators, such as PYTHIA~\cite{Sjostrand:2006za}, PHOJET~\cite{Engel:1995sb}, or HIJING~\cite{Wang:1991hta}
do not include the production of \mbox{(anti-)}nuclei other than
\mbox{(anti-)}protons
and \mbox{(anti-)}neutrons. Therefore, 
nuclei are explicitly injected into underlying PYTHIA (in case of
pp) and HIJING (in case of Pb--Pb) events with a flat momentum distribution. In the
next step, the particles are propagated through the ALICE detector
geometry with the GEANT3 transport code~\cite{Geant:1994zzo}. 
GEANT3 includes a basic description of the interaction of nuclei with the
detector, however, this description is imperfect due to the limited
data available on collisions of light nuclei with heavier materials.
Due to the unknown interaction of anti-nuclei with material, these processes are not included for
anti-nuclei heavier than anti-protons. 
In order to account for these effects, a full detector simulation with GEANT4 as a transport code~\cite{Agostinelli:2002hh,Uzhinsky:2011zz} was used.
Following the approach described in~\cite{Abbas:2013rua}, the correction for interaction of \mbox{(anti-)}nuclei with the detector material 
from GEANT3 was scaled to match the expected values from GEANT4.
An alternative implementation to correct for this effect 
and the relevant uncertainties related to these corrections are discussed in Section~\ref{secAnnihilation}. The
acceptance$\times$efficiency is then obtained as the ratio of the
number of particles detected by the detector 
to the number of generated particles within the
relevant phase space. 

\begin{figure}
\begin{center}
 \includegraphics[width=5.2cm]{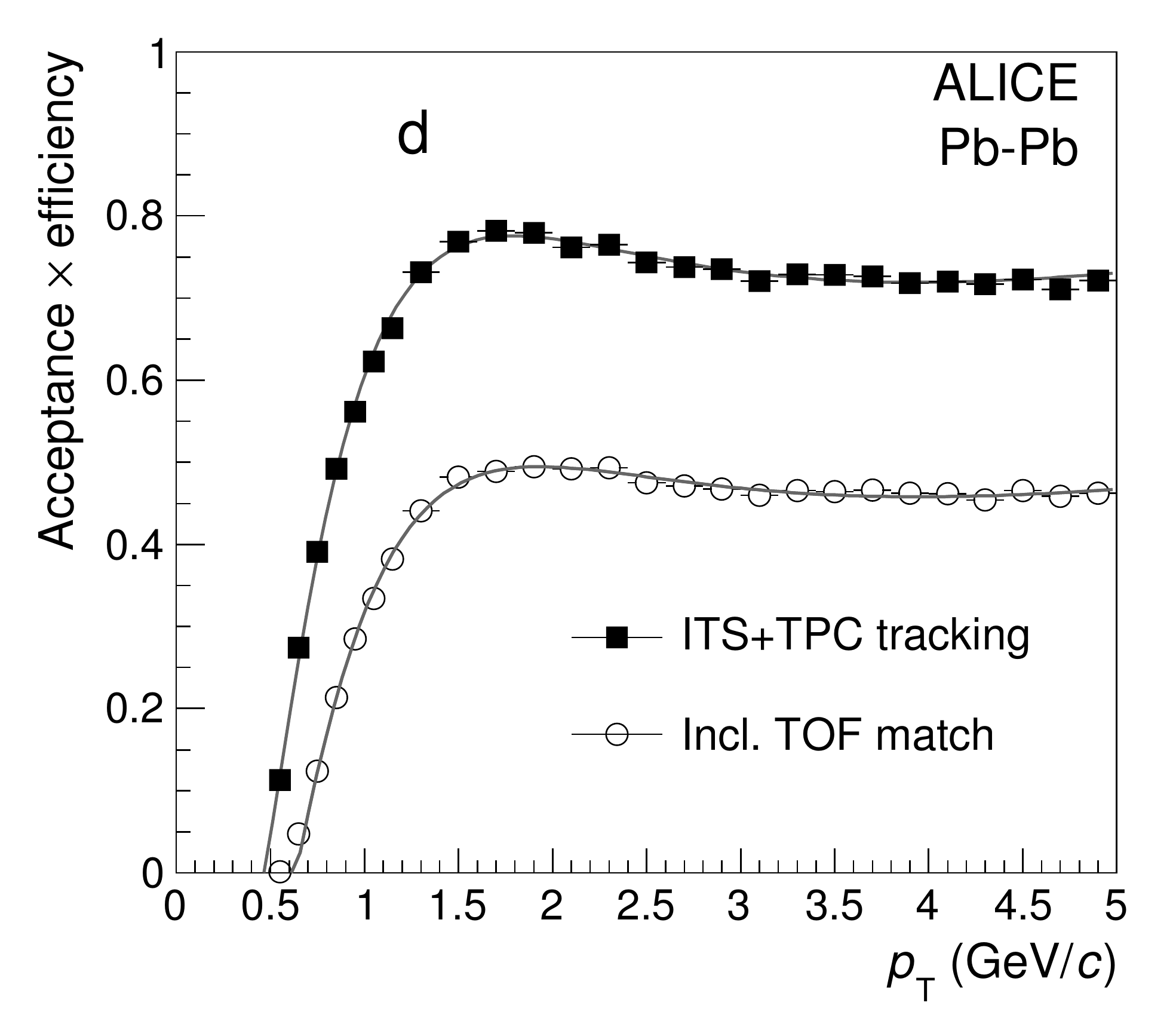}
 \includegraphics[width=5.2cm]{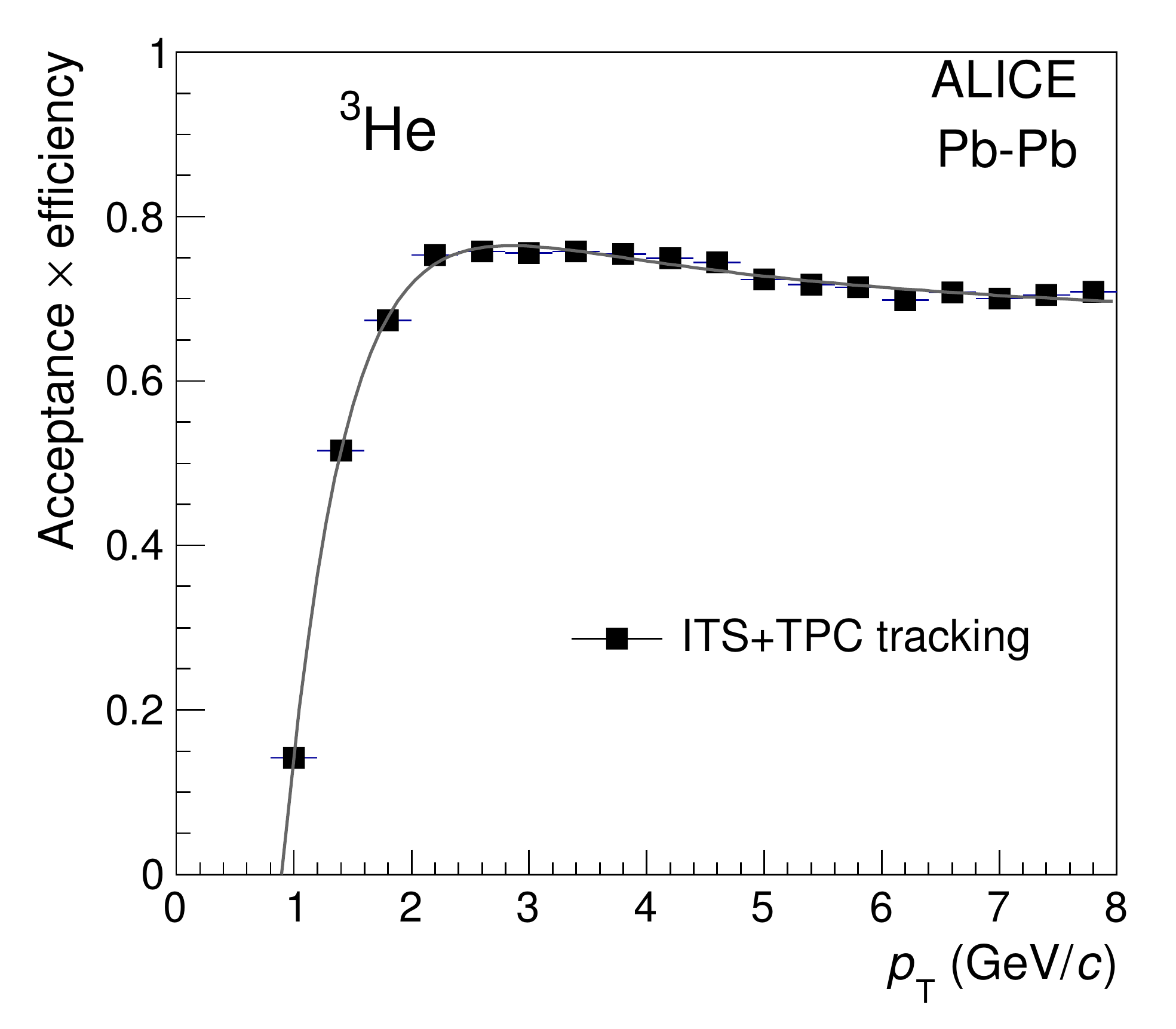}
 \includegraphics[width=5.2cm]{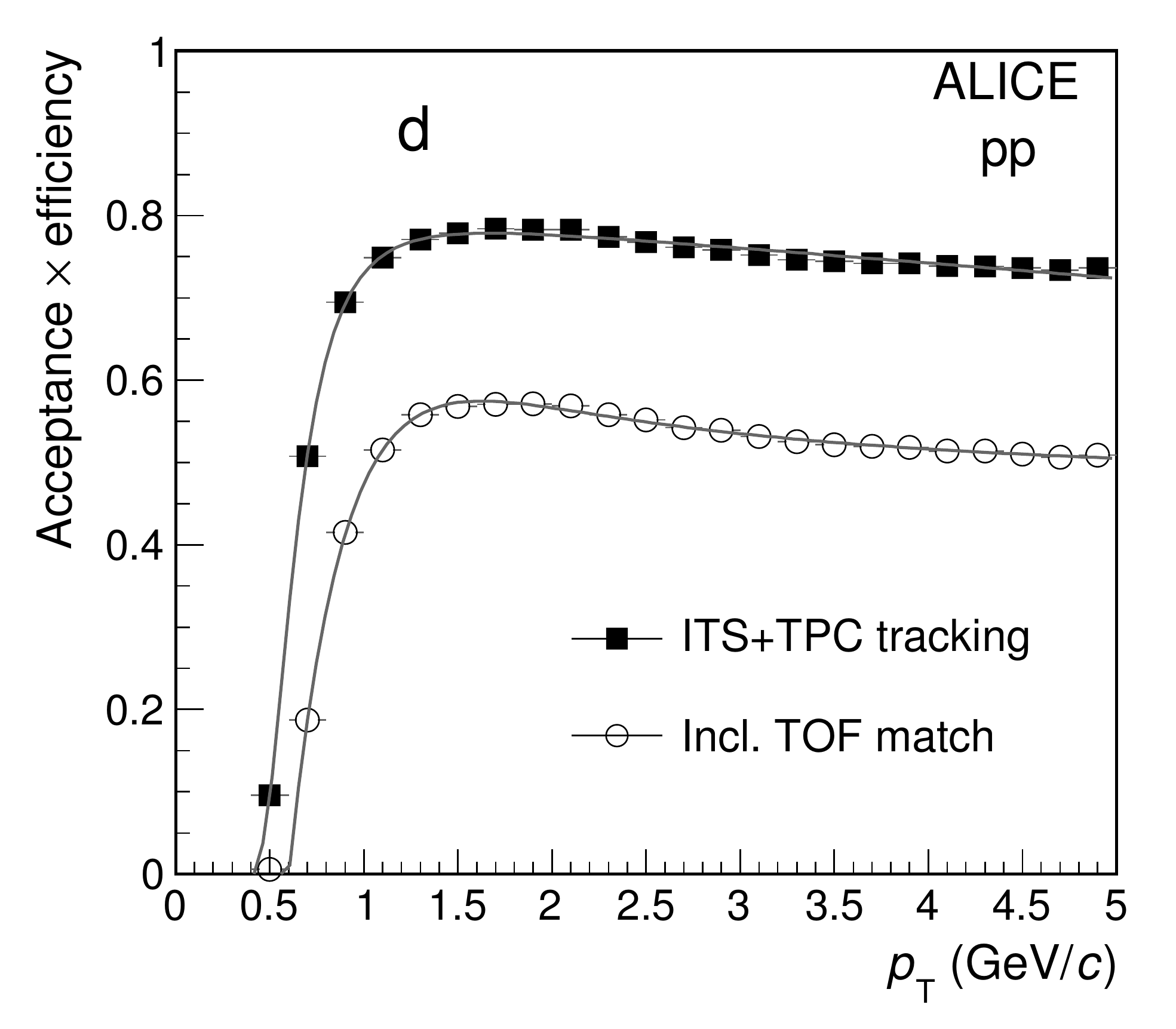}
\caption{Acceptance$\times$efficiency as a function of transverse
momentum (\pt) for deuterons~(left) and for $\rm ^{3}He$ (middle)
in Pb--Pb collisions at \snn\ =~2.76~TeV, as well as for deuterons
in pp collisions at $\sqrt{s}$ =~7~TeV~(right panel). The curves represent a
fit with the function presented in Eq.~(\ref{EffFit}) (see text for details).}
\label{Eff_He3_PbPb}
\end{center}
\end{figure}

Figure~\ref{Eff_He3_PbPb} shows the acceptance$\times$efficiency
for deuterons~(left) and $\rm ^3He$ (middle) as a function of \pt\
for Pb--Pb collisions at \snn\ = 2.76 TeV. In both cases, the
rapid rise of the  efficiency at low \pt\ is determined by energy
loss and multiple scattering processes of the incident particle
with the detector material. The values reach a maximum when the
energy loss becomes smaller and when the track curvature is still
sufficiently large so that a track can cross the dead area between two
TPC readout chambers in a relatively small distance such that the
two track parts can still be connected. For straighter
tracks at higher \pt\ which cross the insensitive region between
two chambers this distance is larger and the connection becomes
more difficult. Thus a slight reduction of the efficiency is
observed until a saturation value is reached. The figure also
shows the lower efficiency values (open points) when in addition a deuteron
track is matched to a hit in the TOF detector. The drop is mainly
caused by the energy loss and multiple scattering in the material
between the TPC and the TOF, by the TOF dead zones corresponding
to other detectors or structures, and by the number of active TOF
channels. The curves represent fits with the empirical functional
form

\begin{equation}
f(p_{\rm T}) = a_{0}~e^{(-a_{1}/p_{\rm T})^{a_{2}}} + a_{3}~p_{\rm T}\; .
\label{EffFit}
\end{equation}

\noindent Here, $a_{0}, a_{1}, a_{2}$, and $a_{3}$ are free
parameters. 
Correcting
the raw spectra with either the fit function or the actual
histogram is found to result in negligible differences with
respect to the total systematic error.

Figure~\ref{Eff_He3_PbPb} (right) also shows 
acceptance$\times$efficiency for the deuterons as a function of \pt\ for pp collisions at $\sqrt{s}$ =
7 TeV. The curve is a fit using the same functional form as used
for the Pb--Pb collisions discussed above. The efficiency has a
similar \pt-dependence as the one for Pb--Pb collisions at \snn\
=~2.76~TeV.
The observed differences are due to variations in the number of active detector
components, mainly in the SPD, for the two data sets.

\subsection{Momentum correction}

\begin{figure}
  \begin{center}
 \includegraphics[width=8.5cm]{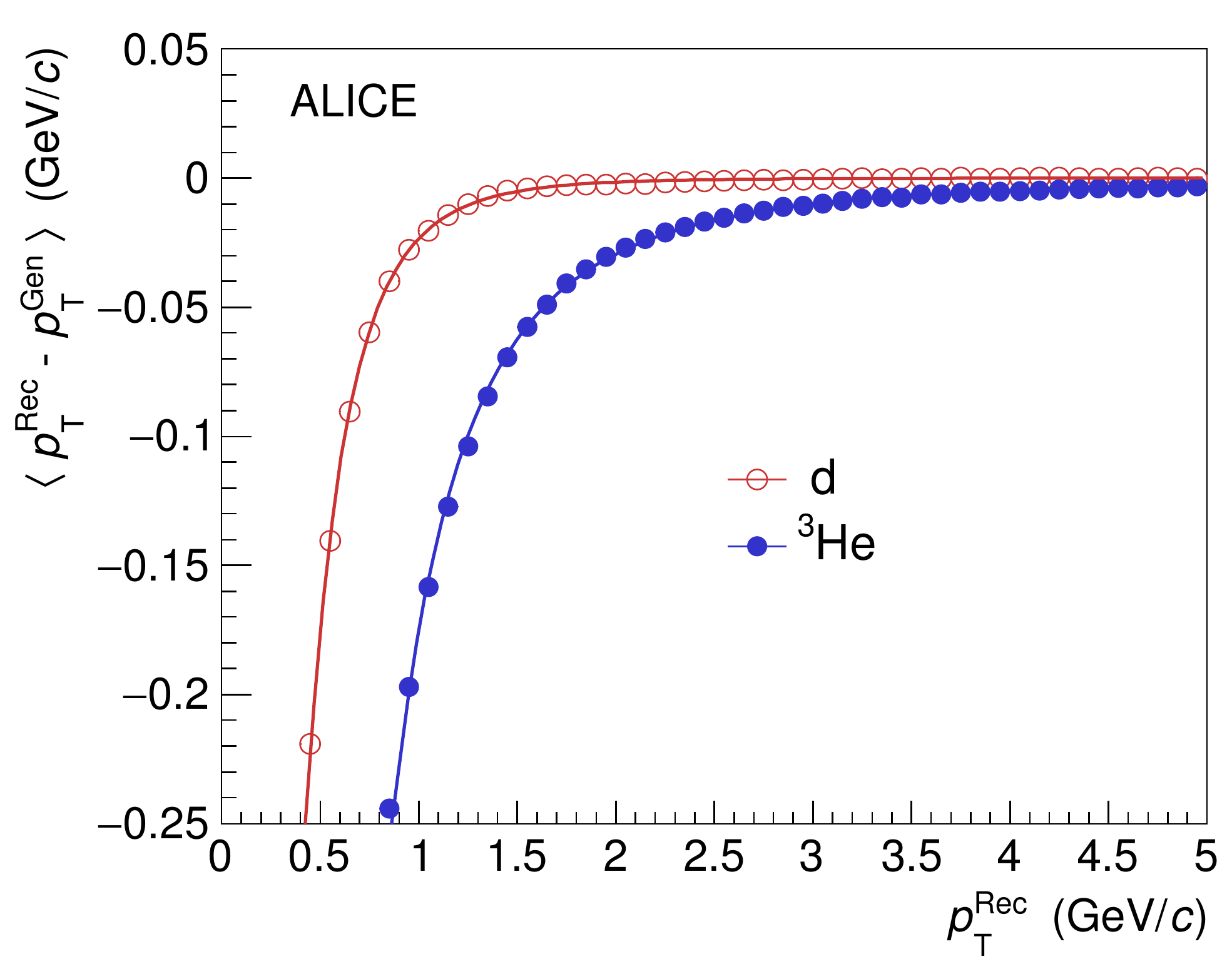}
\caption{(Color online) The average difference between the reconstructed and the
  generated \pt~is plotted as a function of the reconstructed \pt~for simulated
deuterons and $\rm ^3He$ for Pb--Pb collisions at \snn\ = 2.76 TeV.
The lines represent a fit with the functional form as shown in Eq.~(\ref{MomCorFunc}) (see text for details).}
\label{MomCor}
\end{center}
\end{figure}

Low-momentum particles lose  a considerable amount of energy while traversing the detector
material. The track reconstruction algorithm takes into account
the Coulomb scattering and energy loss, assuming the pion mass for
each particle. Therefore, a track-by-track correction for the
energy loss of heavier particles (d/$\rm \bar{d}$
and $\rm ^3He$/$\rm ^3\overline{He}$) is needed. This
correction is obtained from MC simulations, in which the
difference of the reconstructed and the generated transverse
momentum is studied on a track-by-track basis. Figure~\ref{MomCor}
shows the average \pt-difference as a function of the  reconstructed track
momentum ($p_{\mathrm T}^{\rm Rec}$) for deuterons and $\rm ^3He$.
The lines represent the empirical function

\begin{equation}
f(p_{\rm T}) = A + B\left(1 + \frac{C}{p_{\rm T}^{2}}\right)^{D},
\label{MomCorFunc}
\end{equation}

\noindent where the free parameters $A, B, C,$ and $D$ are
extracted from a fit.
It can be seen that the correction becomes largest for the heaviest particles 
at low momenta. This reflects the typical 1/$\beta^{2}$-behavior of the energy loss.
The difference in transverse momentum is corrected on a track-by-track basis in the analysis. 
This energy loss correction has been applied both for pp and for Pb--Pb
collisions. The same correction in rapidity has also been studied and
found to result in negligible changes in the final spectra.

\subsection{Systematic uncertainties}

Individual contributions to the systematic error of the
measurement are summarized in Table~\ref{tab.:SystCentral} and are
discussed in detail in the following. The systematic uncertainty
related to the identification of the nuclei is smaller in the
\pt-region in which the energy loss in the TPC provides a clear
separation compared to those in which the identification is mainly
based on the TOF information. The error is of the order
of 1\% for deuterons at low momenta and for the full \pt-range
studied for $^{3}$He-nuclei. In the TOF part (\pt\ $>$ 1.4~GeV/$c$) 
of the deuteron spectrum, the error is considerably larger due to the presence of background
and has been estimated as 5\% on the basis of different signal extraction
methods: the raw yields obtained from the signal fit and from bin
counting are compared. 
The estimates of the uncertainties related to the tracking and matching are
based on a variation of the track cuts
and are found to be less than 4\% and independent of
the particle species. In addition to this, a variation in the
momentum correction leads to differences of similar magnitude at
lower momenta and are added in quadrature.

Contamination from secondaries originating from interactions of
primary particles with the detector material dominates the
systematic error at low transverse momenta, but it decreases
exponentially towards higher momenta. These uncertainties are
estimated by a variation of the fit range and templates. Their values
amount to about 20\% in the lowest \pt-bin for deuterons and for
\he\ in most central events. For all other centralities and
transverse momentum regions, it is significantly lower. Feed down
from weakly decaying hyper-nuclei is negligible for deuterons. The
only relevant decay of the hyper-triton, 
$^{3}_{\Lambda}{\mathrm{H}} \rightarrow \mathrm{d} + \mathrm{p} +
\mathrm{\pi}^{-}$, results in a negligible contamination, because of 
the roughly 700 times smaller production cross section of the hyper-triton
with respect to the deuteron \cite{Andronic:2010qu,Cleymans:2011pe}. On the other hand, the decay
$^{3}_{\Lambda}{\mathrm{H}} \rightarrow {\mathrm{^{3}He}} +
\mathrm{\pi}^{-}$ contaminates the \he-spectrum as these particles
are produced with similar abundance. This background is
conceptually similar to the feed down of $\Lambda$ decays into the
proton-spectrum \cite{Abelev:2013xaa}  though the relevant
branching ratio in the case of $^{3}_{\Lambda}{\mathrm{H}}$
(25\%)~\cite{Kamada:1997rv}
is assumed 
to be considerably lower than in the case of $\Lambda$ (64\%). A detailed MC study shows
that only about 4-8\% of all $^{3}_{\Lambda}{\mathrm{H}}$ decaying
into \he\ pass the track selection criteria of primary \he.
Therefore, the remaining contamination has not been subtracted and the
uncertainty related to it was further investigated by a
variation of the DCA$_{xy}$-cut in data and a final error of about 5\% is assigned.
Uncertainties in the material budget
have been studied by simulating events varying the amount of material by $\pm10\%$. This leads to variations in the
efficiency of about 5\% in the lowest \pt-bins.
The hadronic interaction of nuclei with the detector material gives rise to an additional
uncertainty of about 6\% for deuteron and for \he. The material between TPC and TOF needs
to be considered only for the deuteron spectrum above \pt\ $>$ 1.4~GeV/$c$ and increases the uncertainty
by additional 7\%. The corresponding corrections
for anti-nuclei are significantly larger and less precisely determined
because of the missing knowledge of the relevant elastic and inelastic
cross sections.
Details of the systematics originating from differences between the available models
are discussed in the next section.

In general, the individual contributions to the systematic error do not show a significant dependence on the event multiplicity.
The only exception is given by the uncertainty of the correction for secondaries from material, which changes from about 20\% in central to about 4\% in peripheral Pb--Pb or pp collisions, respectively. All other contributions are found to be independent of event multiplicity.

\begin{table*}
 \begin{center}
  \vspace{0.20cm}
  \begin{tabular}{c|c c|c c}
  \hline
  \hline
  Source                 &  \multicolumn{2}{c|}{d}          &\multicolumn{2}{c}{$^{3}$He}   \\
  \hline
                         & 0.7 GeV/$c$ & 4 GeV/$c$   & 2 GeV/$c$ & 8 GeV/$c$   \\
  \hline
  PID                    & 1\% & 5\%                 & 1\% &  1\%                \\
  Tracking and matching  & 6\% &  4\%                 &  6\%    &     4\%              \\

  Secondaries material   &20\%  &   1\%                &  20\%  &    1\%            \\
  Secondaries weak decay & \multicolumn{2}{c|}{negl.}       &  \multicolumn{2}{c}{5\% }    \\

  Material budget        & 5\%     &    1\%                &  3\%    &       1\%            \\
  Hadronic interaction   & \multicolumn{2}{c|}{6\%}         &  \multicolumn{2}{c}{6\%}  \\

  \hline
  \hline
  \end{tabular}
  \caption{\label{tab.:SystCentral} Summary of the main contributions to the systematic uncertainties. See text for details.}
 \end{center}
\end{table*}

\section{\label{secResults} Results}

\begin{figure*}
\begin{center}
\includegraphics[height=5.2cm]{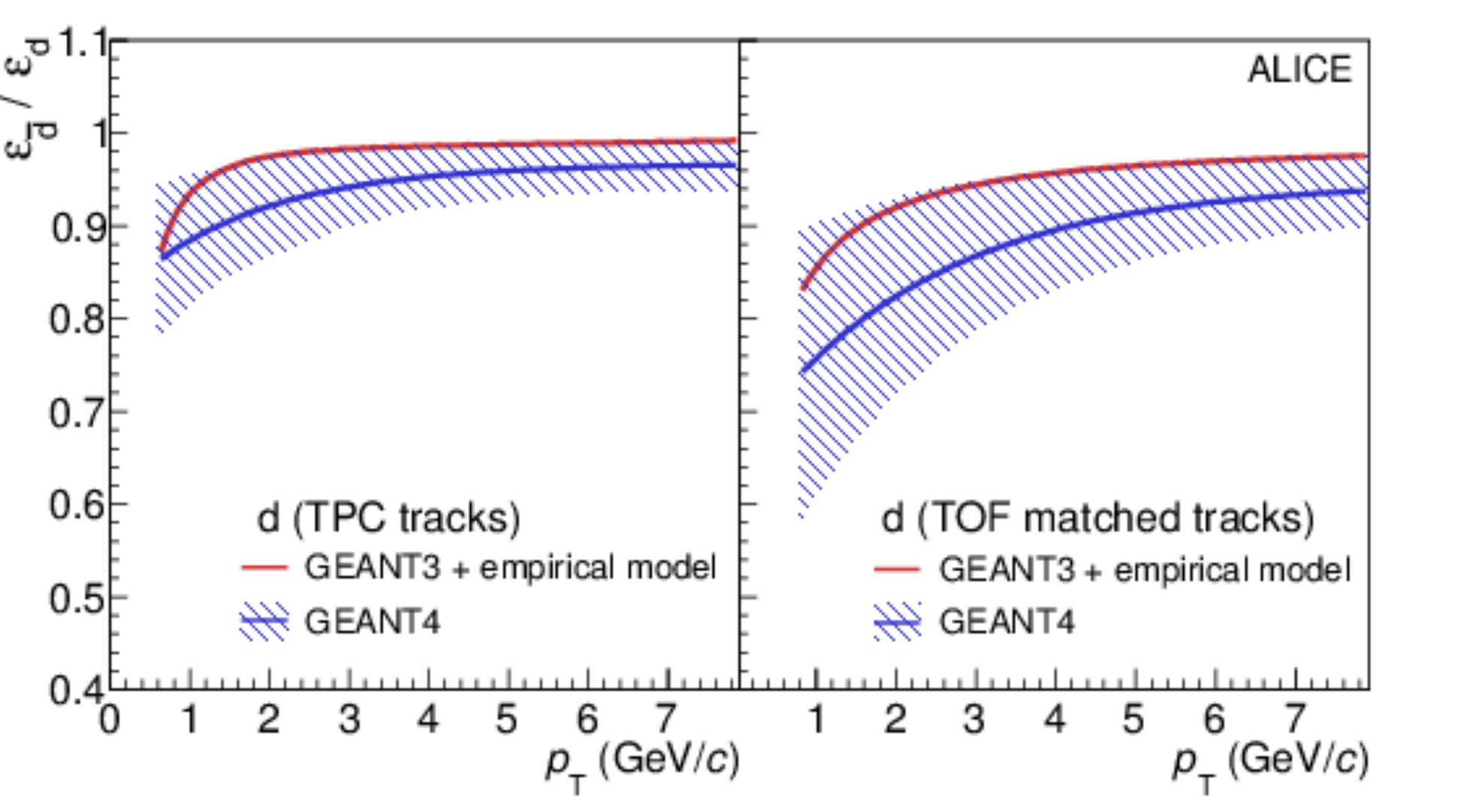} 
 \includegraphics[height=5.2cm]{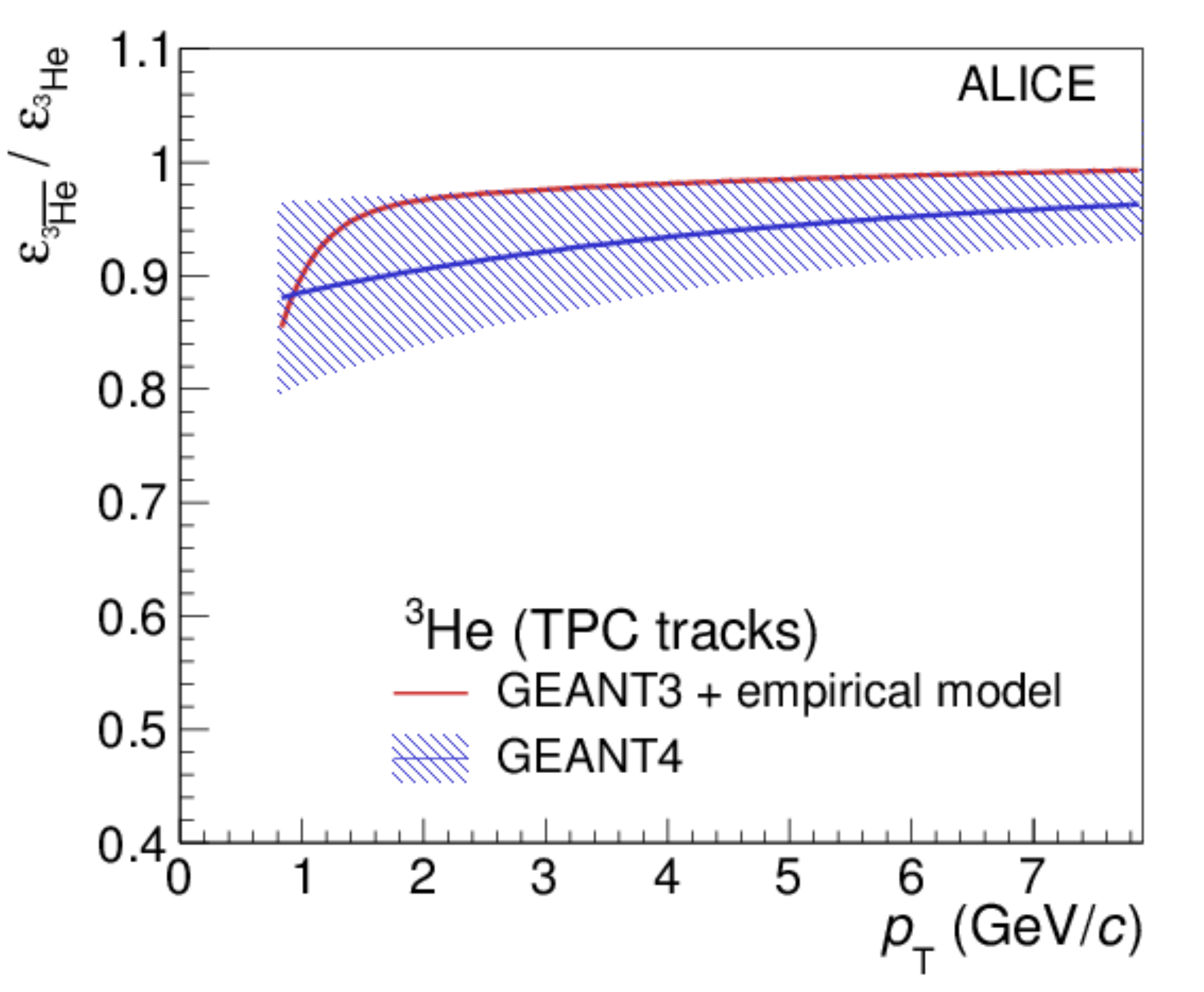}
\caption{(Color online) Ratio of anti-particle to particle efficiency 
  based on GEANT4 and a modified version of GEANT3 including an
  empirical model to describe the hadronic interaction of anti-nuclei
  for \mbox{(anti-)}deuterons (left) and for  \mbox{(anti-)}\he\
  (right). The estimate of the systematic uncertainty for the hadronic interaction based on the difference between the two models is indicated by the blue band.}
 \label{figAbsorption}
 \end{center}

\begin{center}
 \includegraphics[width=10.0cm]{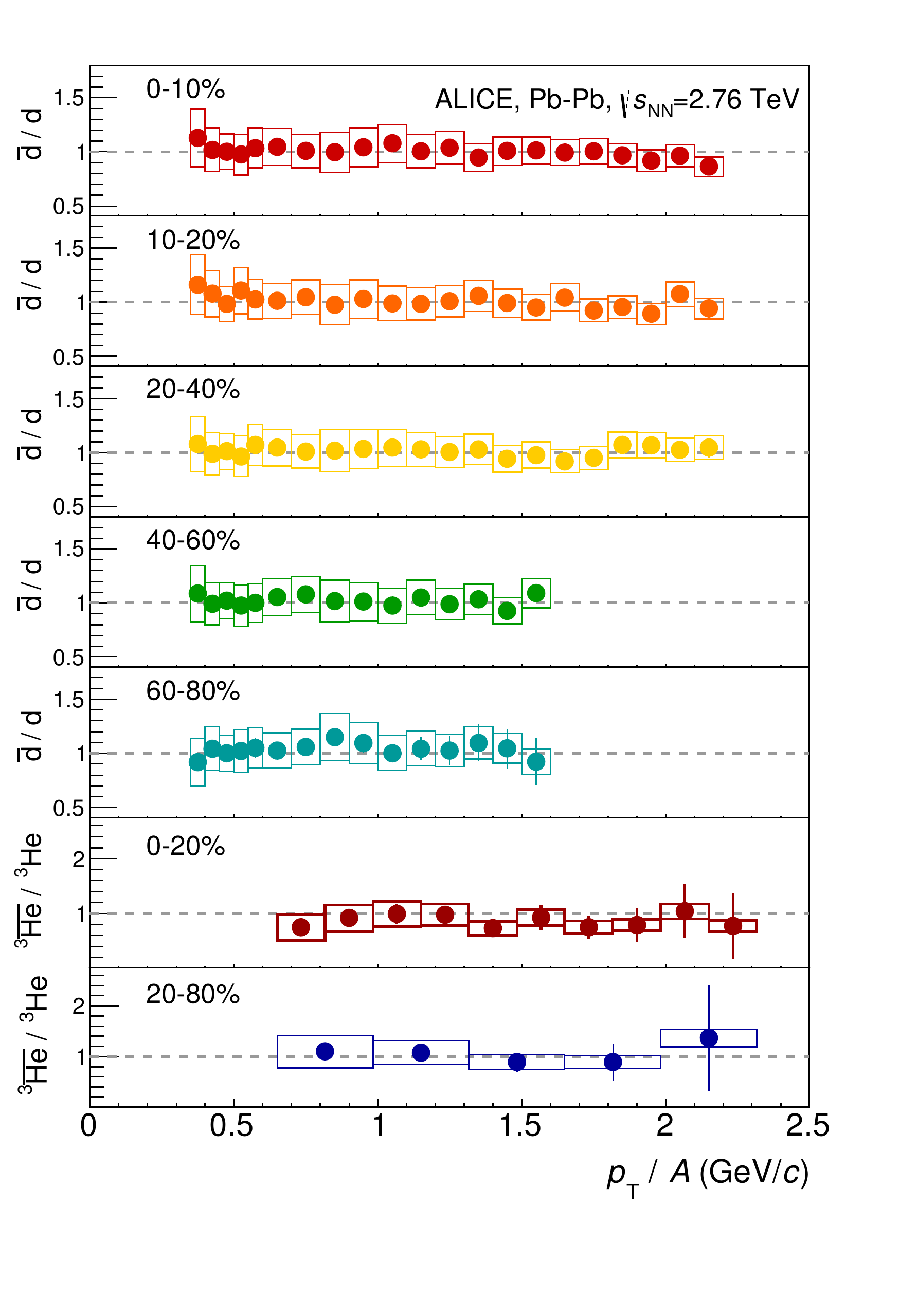} 
 \caption{(Color online) Ratios of $\overline{\rm d}$ and ${\rm d}$ as well as of $^{3}\overline{\rm He}$ and \he~versus \pt\ per nucleon for various
  centrality classes in Pb--Pb collisions at \snn\ = 2.76 TeV. Boxes describe the systematic uncertainties, vertical lines the statistical ones.}
  \label{figAntiParticleToParticleRatios}
 \end{center}
\end{figure*}

\subsection{\label{secAnnihilation}Anti-particle to particle ratios and hadronic interaction of anti-nuclei}

For a measurement of particle to anti-particle ratios, the
correction of the hadronic interaction of the 
emitted particles with the
detector material has to be precisely known.
The relevant cross sections are only poorly measured for
anti-nuclei heavier than \pbar. The only available data for
anti-deuterons from the U-70 Serpukhov
accelerator~\cite{Denisov:1971im,Binon:1970yu} are measured at relatively high
momenta ($p$~=~13.3~GeV/$c$ and $p$~=~25.0~GeV/$c$) and provide only a rough constraint.
Two approaches are considered to model the correction for hadronic interaction.
Firstly, the anti-nuclei cross sections are approximated in a
simplified empirical model by a combination of the anti-proton
($\sigma_{\bar{\rm p},A}$) and anti-neutron ($\sigma_{\bar{\rm
n},A}$) cross sections. Following the approach presented in~\cite{Moiseev}, 
the cross section $\sigma_{\bar{\rm d},A}$ for an anti-deuteron 
on a target material with mass number $A$ is then e.g. given by

\begin{equation}
\sigma_{\bar{\rm d},A} = \bigl( \sigma_{\bar{\rm p},A}^{3/2} +
\sigma_{\bar{\rm n},A}^{3/2} \bigr)^{2/3} \, K(A) \;,
\end{equation}

\noindent where the scaling factor $K(A)$ is determined from the
same procedure applied to the measured inelastic cross sections of nuclei
and protons. Details of the method can be found in~\cite{Moiseev}.
This approach is implemented as a modification to
GEANT3. However, it does not account for elastic scattering processes and
is therefore only used  for the estimation of the systematic uncertainty. 
Secondly, the anti-nucleus--nucleus cross sections are
determined in a more sophisticated model with Glauber calculations
based on the well-measured total and elastic p\pbar\ cross section
\cite{Uzhinsky:2011zz}. It is implemented in the GEANT4 software
package \cite{Agostinelli:2002hh}.

The relevant correction factor for the anti-particle to particle ratio is given by the ratio of the efficiencies
in which all effects cancel except of those related to the hadronic interaction with the detector material. 
The efficiency ratios for anti-deuterons and for $^{3}\mathrm{\overline{He}}$ nuclei using the two models described above
(modified GEANT3 and GEANT4) are shown in Fig.~\ref{figAbsorption}. The applied correction factors
are parameterized with the same function which was used for a similar
study in \cite{Abbas:2013rua}. The absorption correction is larger for tracks which
are required to reach the TOF detector due to the additional
material behind the TPC, mainly the support structure and the Transition
Radiation Detector (TRD). In the following, results corrected with
GEANT4 are presented. Based on the discrepancy between the two models,
an uncertainty of 60\% of the difference between
the efficiency for particles and anti-particles is assumed for the
absorption correction. It is indicated by the blue band in Fig.~\ref{figAbsorption}.

\begin{table*}
 \begin{center}
  \vspace{0.20cm}
 \begin{tabular}{c|c|c}
 \hline
  \hline
 Anti-nuclei/nuclei & Centrality   &  Ratio \\ 
   \hline
 \multirow{5}{*}{ $\bar{\rm d}$/d} &   0-10\% &  0.98 $\pm$ 0.01 $\pm$0.13 \\
   & 10-20\% &    0.99 $\pm$ 0.01  $\pm$ 0.13  \\
   & 20-40\% &   1.01 $\pm$ 0.01 $\pm$ 0.14 \\
   & 40-60\% & 1.02  $\pm$ 0.01  $\pm$ 0.16 \\
   & 60-80\% &  1.02 $\pm$ 0.02  $\pm$ 0.16 \\
  \hline
 \multirow{2}{*}{$^{3}\overline{\rm He}$/$^{3}$He} & 0-20\% & 0.83 $\pm$ 0.08 $\pm$  0.16 \\
    & 20-80\% & 1.03 $\pm$ 0.14 $\pm$ 0.18 \\
  \hline
  \hline
  \end{tabular}
  \caption{\label{tab.:RatioTable} Anti-particle to particle ratios
    for various centrality classes in Pb--Pb collisions at \snn\ =
    2.76 TeV. The first error represents the statistical error and the second one is the systematic error. See text for details.}
 \end{center}
\end{table*}

Applying this model-based correction to the data, 
leads to
\dbar/d and $^{3}\overline{\rm He}$/\he\ ratio shown in Fig.~\ref{figAntiParticleToParticleRatios} for various centrality bins in Pb--Pb collisions. 
Both ratios are consistent with unity and exhibit a constant behavior as a function of \pt\ 
as well as of collision centrality. Since the same statements hold true for the $\bar{\rm{p}}/\rm{p}$ ratios~\cite{Abelev:2013vea},  
these observations are in agreement with expectations from the  thermal-statistical and coalescence models \cite{Cleymans:2011pe}
which predict a ratio of \dbar/d~=~$(\bar{\rm{p}}/\rm{p})^{2}$ and $^{3}\overline{\rm He}$/\he~=~$(\bar{\rm{p}}/\rm{p})^{3}$.
Table~\ref{tab.:RatioTable} show the anti-particle to particle ratios for various centrality
classes in Pb--Pb collisions at \snn\ = 2.76 TeV.

Ongoing studies on the hadronic interaction of anti-nuclei in the material between
the TPC and TOF will allow to constrain the uncertainties of the currently purely model
based corrections and to replace them with data driven ones. As the spectra for
nuclei and anti-nuclei are consistent within the currently large uncertainties, only
the spectra of nuclei are provided in the following.

\subsection{Spectra of nuclei}

\begin{figure}[h]
\begin{center}
 \includegraphics[width=8.5cm]{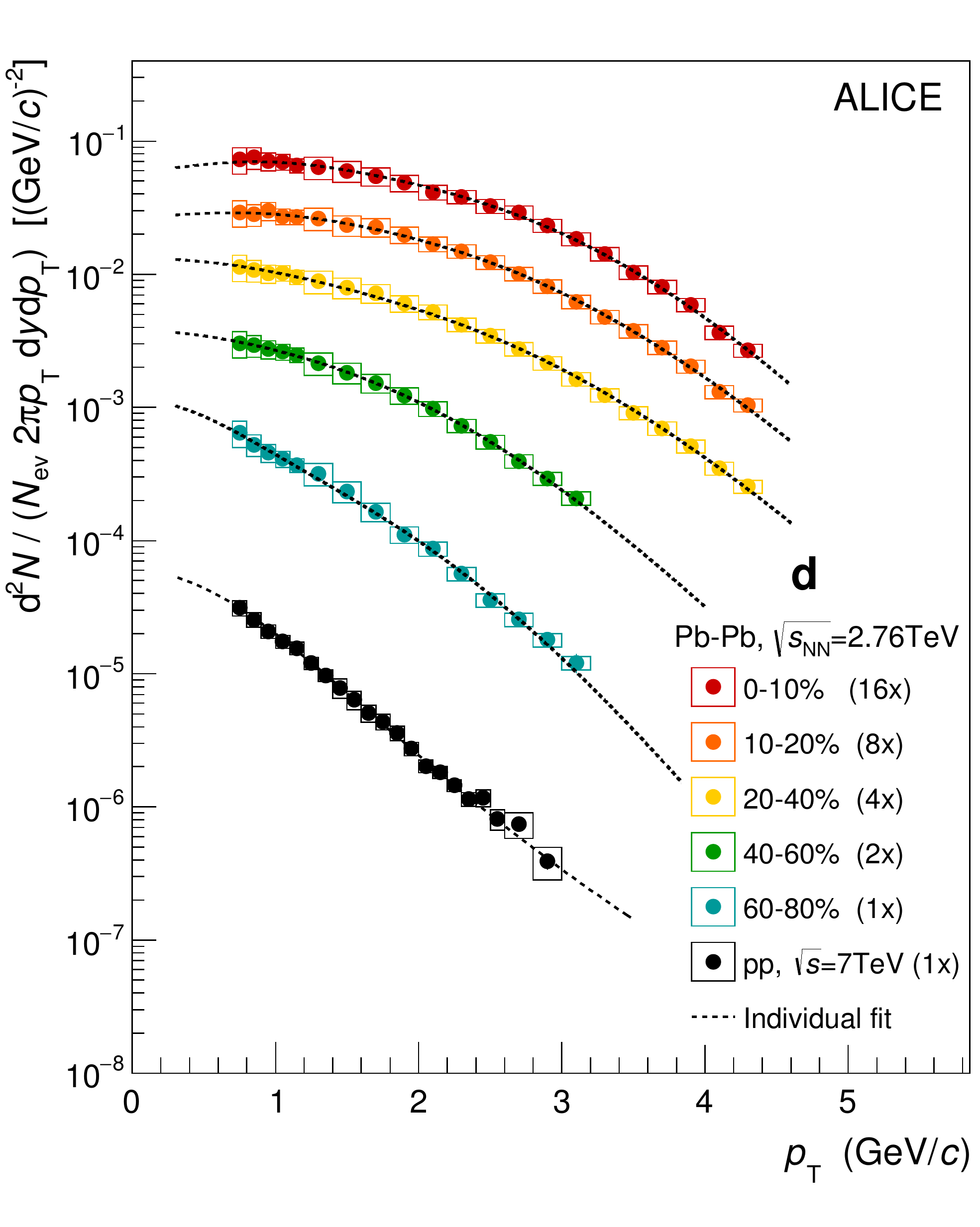}
\caption{(Color online) Efficiency and acceptance corrected deuteron spectra for
Pb--Pb collisions at $\sqrt{s_{\rm NN}}$ = 2.76 TeV in various
centrality classes and for inelastic pp collisions at $\sqrt{s}$ =
7 TeV. The dashed lines represent an individual fit with the BW function (Eq.~\ref{bwfunc}) in the case of Pb--Pb spectra and with the function
presented in Eq.~(\ref{levyfunc}) in the case of the pp spectrum
(see text for details). The boxes show systematic error and
vertical lines show statistical error separately.}
\label{finalspectra}
\end{center}
\end{figure}

The final spectra of deuterons obtained in Pb--Pb and pp
collisions are shown in Fig.~\ref{finalspectra}. The statistical
and systematic errors are shown separately as vertical lines and
boxes, respectively. In pp collisions, the spectrum is normalized
to the number of all inelastic collisions ($N_{\rm{INEL}}$) which includes a correction for
trigger inefficiencies (see \cite{Abelev:2012sea,Abelev:2013ala} for details). 
It is fitted with the following function~\cite{Tsallis:1987eu,Abelev:2006cs,Aamodt:2010my} 
that has been used for lighter particles
\begin{equation}
\frac{1}{2\pi p_{\rm T}} \frac{{\rm d^2}N}{{\rm d}p_{\rm T} {\rm d}y} = 
\frac{{\rm d} N}{{\rm d} y} \frac{(n-1)(n-2)}{2\pi~nC(nC + m_0(n-2))}
\left(1 + \frac{m_{\rm T} - m_0}{nC}\right)^{-n}
\label{levyfunc}
\end{equation}
\noindent with the fit parameters $C$, $n$, and the
\dndy. 
The parameter $m_0$ corresponds to the mass of the particle under
study (deuteron) at rest and $m_{\rm T} =\sqrt{m_{0}^{2} + p_{\rm
T}^{2}}$ to the transverse mass. As in the case of lighter
particles, the function is found to describe the deuteron
\pt\ spectrum well in the measured range with a $\chi^2/{\rm ndf}$ of
 0.26 .
The fit function is used for the extrapolation to the unmeasured region
at low and high transverse momenta 
(about 45\% of the total yield)
and a \pt-integrated yield of 
\dndy\ = $(2.02 \pm 0.34(\rm syst)) \times 10^{-4}$ is obtained.

While statistical errors are negligible, the systematic error is
dominated by the uncertainty related to the extrapolation (13\%)
which is evaluated by a comparison of different fit functions \cite{Adler:2003cb}
(Boltzmann, $m_{\rm{T}}$-exponential, $p_{\rm{T}}$-exponential, Fermi-Dirac, Bose-Einstein).
Based on the same extrapolation in the unmeasured region of the
spectrum, a mean transverse momentum $\langle$\pt$\rangle$ of
$1.10 \pm 0.07$ GeV/$c$ is obtained.

\begin{figure}
\begin{center}
 \includegraphics[width=8.5cm]{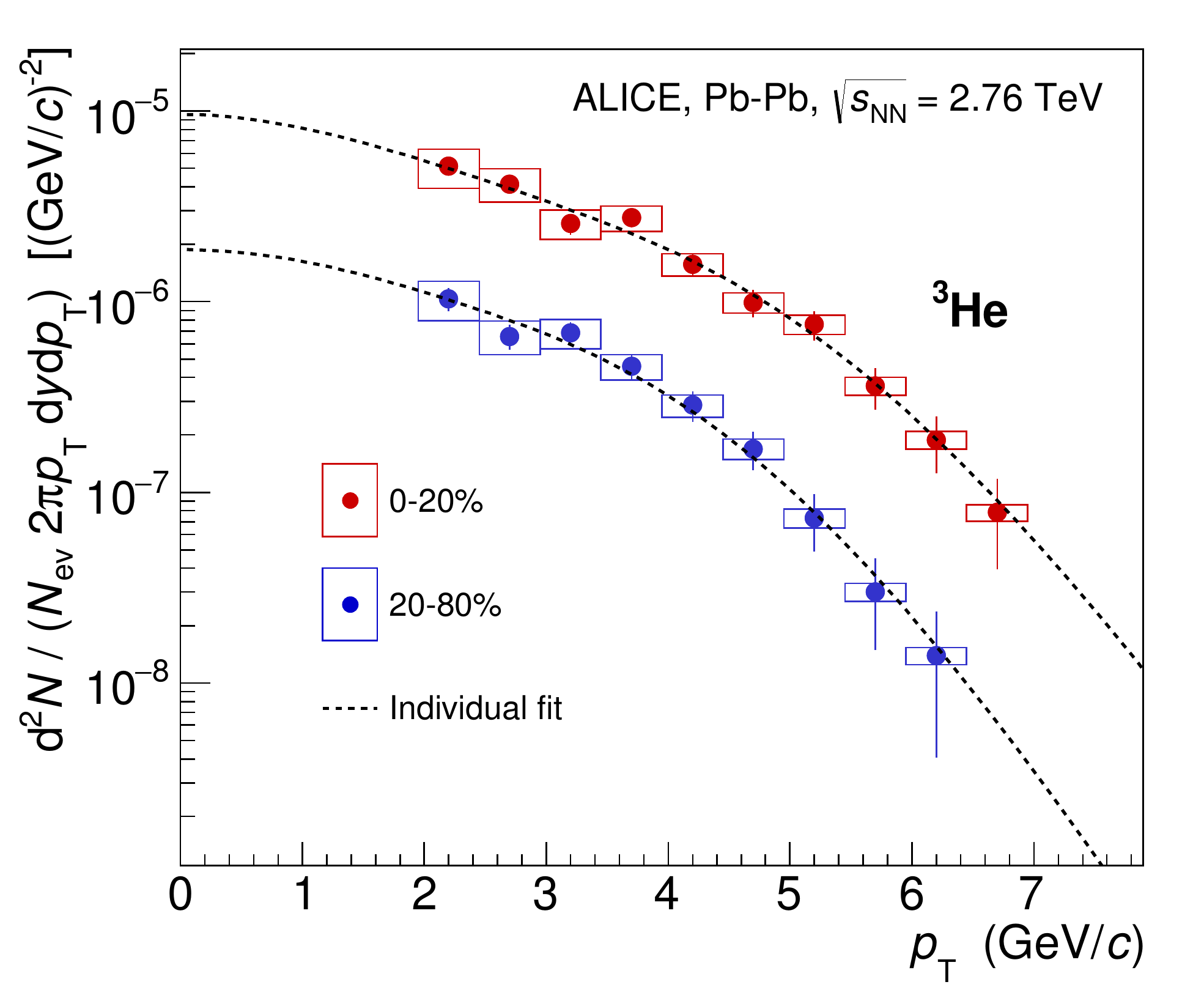}
\caption{(Color online) $^{3}$He spectra for two centrality classes (0--20\% and
  20--80\%) are shown for Pb--Pb collisions at $\sqrt{s_{\rm NN}}$
  = 2.76 TeV. The spectra are fitted individually with the BW
  function (dashed lines). The systematic and statistical errors are shown 
  by boxes and vertical lines, respectively.}
 \label{He3_fit}
\end{center}
\end{figure}

The final spectra of deuterons and $\rm ^{3}He$
for Pb--Pb collisions at $\sqrt{s_{\rm NN}}$ = 2.76 TeV are shown
in Figs.~\ref{finalspectra} and \ref{He3_fit} for various choices
of the collision centrality. Again, the systematic and statistical errors
are shown separately by boxes and vertical lines, respectively.
The \pt\ distributions show a clear evolution, becoming harder as
the multiplicity increases. A similar behavior is observed for protons, which have been 
successfully described by models that incorporate a significant radial flow \cite{Abelev:2013vea}.

\begin{figure}
\begin{center}
\includegraphics[width=8.5cm]{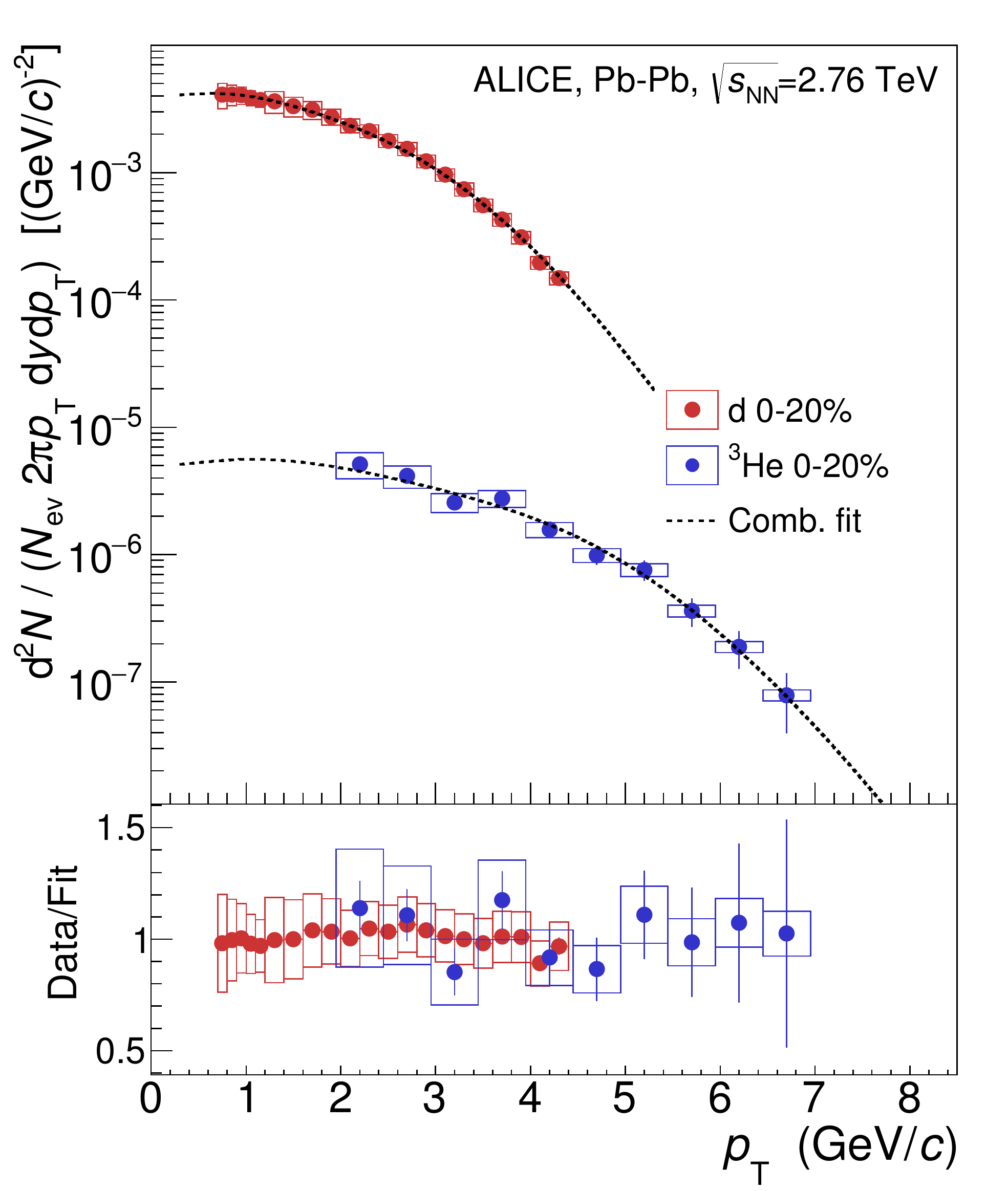}
\caption{(Color online) The top panel shows the combined fit of deuteron and $\rm
  ^3He$ spectra with the BW function for 0--20\% centrality for
Pb--Pb collisions at \snn\ = 2.76 TeV. The systematic and statistical errors are shown 
  by boxes and vertical line, respectively. The lower panel shows the
  deviation of the spectra from the BW fits. }
 \label{d_He3_fit}
\end{center}
\end{figure}

%

The spectra obtained in Pb--Pb collisions are individually fitted
with the blast-wave (BW)
model for the determination of \pt-integrated yields and
$\langle$\pt$\rangle$.
This model~\cite{Schnedermann:1993ws} describes particle
production properties by assuming that the particles are emitted
thermally from an expanding 
source.
The functional form of the model is given by
\begin{equation}
\frac{1}{p_{\rm T}}\frac{\mathrm{d}N}{\mathrm{d}p_{\rm T}} \propto \int_0^Rr
\,\mathrm{d}r \, m_{\rm T} I_0 \Bigl( \frac{p_{\rm T} 
\sinh\rho}{T_{\rm kin}} \Bigr) K_1 \Bigl(\frac{m_{\rm T} \cosh\rho}{T_{\rm
kin}} \Bigr) \; , \label{bwfunc}
\end{equation}
\noindent where the velocity profile $\rho$ is described by

\begin{equation}
\rho = \tanh^{-1}\beta = \tanh^{-1} \Bigl( \beta_{\rm S}(r/R)^n
\Bigr) \;.
\end{equation}

\noindent Here $I_{0}$ and $K_{1}$ are the modified Bessel
functions, $r$ is the radial distance from the center of the
fireball in the transverse plane, $R$ is the radius of the
fireball, $\beta(r)$ is the transverse expansion velocity,
$\beta_{\rm S}$ is the transverse expansion velocity at the
surface, $n$ is the exponent of the velocity profile, and $T_{\rm
kin}$ is the kinetic freeze-out temperature. The free parameters
in the fit are $T_{\rm kin}$, $\beta_{\rm S}$, $n$, and a
normalization parameter. Here, we present two 
alternatives: fitting the
two particles separately (Figs.~\ref{finalspectra} and
\ref{He3_fit}) and simultaneously (Fig.~\ref{d_He3_fit}). The
extracted values of the kinetic freeze-out temperature and radial
flow velocity are discussed in more detail in the next section.
The results of these fits are summarized in
Table~\ref{tab.:summary_values}, where the values of d$N$/d$y$ and
$\langle p_{\rm T} \rangle$ are also reported. The d$N$/d$y$
values are extracted by individually fitting the spectra with the
BW model. The extrapolation to \pt~$= 0$ introduces an additional
error which is again evaluated by a comparison of different fit
functions and amounts to about 6\% for central and 13\% for
peripheral collisions for the deuteron yields. In the \he~ case,
it contributes about 17\% and 16\% to the total systematic errors
for the 0--20\% and 20-80\% centrality class, respectively.

\begin{table*} 
\begin{center}
\resizebox{\linewidth}{!}{ 
\begin{tabular}{lccccccc}
\hline
\hline
\noalign{\smallskip}
Centrality         &$\langle \beta \rangle$ &  $T_{\rm kin}$ (MeV)  &
$n$                &  d$N$/d$y$                                  &
            & $\langle\ p_{\rm T}\ \rangle$ (GeV/$c$) & $\chi^2$/ndf \\

\hline\noalign{\smallskip}

d (0-10\%)     &   $0.630 \pm 0.003 $   &   $77 \pm 2$    &   $0.75 \pm 0.05 $  &   $ (9.82 \pm 0.04 \pm 1.58) \times 10^{-2} $ &  & $2.12 \pm 0.00 \pm 0.09$  & 0.10  \\
d (10-20\%)   &   $0.613 \pm 0.004 $   &   $96 \pm 2$    &   $0.78 \pm 0.06 $  &   $ (7.60 \pm 0.04 \pm 1.25) \times 10^{-2} $ &  & $2.07 \pm 0.01 \pm 0.10$  & 0.07  \\
d (20-40\%)   &   $0.572 \pm 0.004 $   &   $100 \pm 2$  &   $0.96 \pm 0.07 $  &   $ (4.76 \pm 0.02 \pm 0.82) \times 10^{-2} $ &  & $1.92 \pm 0.00 \pm 0.11$  & 0.07  \\
d (40-60\%)   &   $0.504 \pm 0.017 $   &   $124 \pm 7$  &   $1.04 \pm 0.19 $  &   $ (1.90 \pm 0.01 \pm 0.41) \times 10^{-2} $ &  & $1.63 \pm 0.01 \pm 0.09$  & 0.01  \\
d (60-80\%)   &   $0.380 \pm 0.010 $   &   $108 \pm 3$  &   $1.85 \pm 0.35 $  &   $ (0.51 \pm 0.01 \pm 0.14) \times 10^{-2} $ &  & $1.29 \pm 0.01 \pm 0.14$  & 0.21  \\

\hline 

$^{3}$He (0-20\%)  & $0.572  \pm 0.006$      & $101 \pm 61$ & $1.02  \pm 0.02$	&	  $ (2.76 \pm 0.09 \pm 0.62) \times 10^{-4} $ &  	&	 $2.83 \pm 0.05 \pm 0.45$ & 0.49 \\
$^{3}$He (20-80\%) & $0.557 \pm 0.007$      & $101 \pm 37$ & $0.99 \pm 0.03$  	&	  $ (5.09 \pm 0.24 \pm 1.36) \times 10^{-5} $ &  	&	 $2.65 \pm 0.06 \pm 0.45$  & 0.20 \\

\hline
d, $^{3}$He (0-20\%)  & $0.617 \pm0.009$  & $83 \pm 22$ &  $0.81 \pm 0.06$ &             &      &   &  0.32 \\
\hline
\hline
\end{tabular}
}
\caption{\label{tab.:summary_values}
 Summary of extracted yields d$N$/d$y$ and mean transverse
momenta $\langle p_{\rm T} \rangle$ based on the BW 
individual fits performed on the spectra for Pb--Pb collisions at
\snn\ = 2.76 TeV. The first error on \dndy~and $\langle p_{\rm T} \rangle$ represents
the statistical error and the second one is the combination of
systematic and extrapolation errors, added in quadrature.
See text for details.
}
\end{center}
\end{table*} 

\begin{figure}[h]
\begin{center}
 \includegraphics[width=8.5cm]{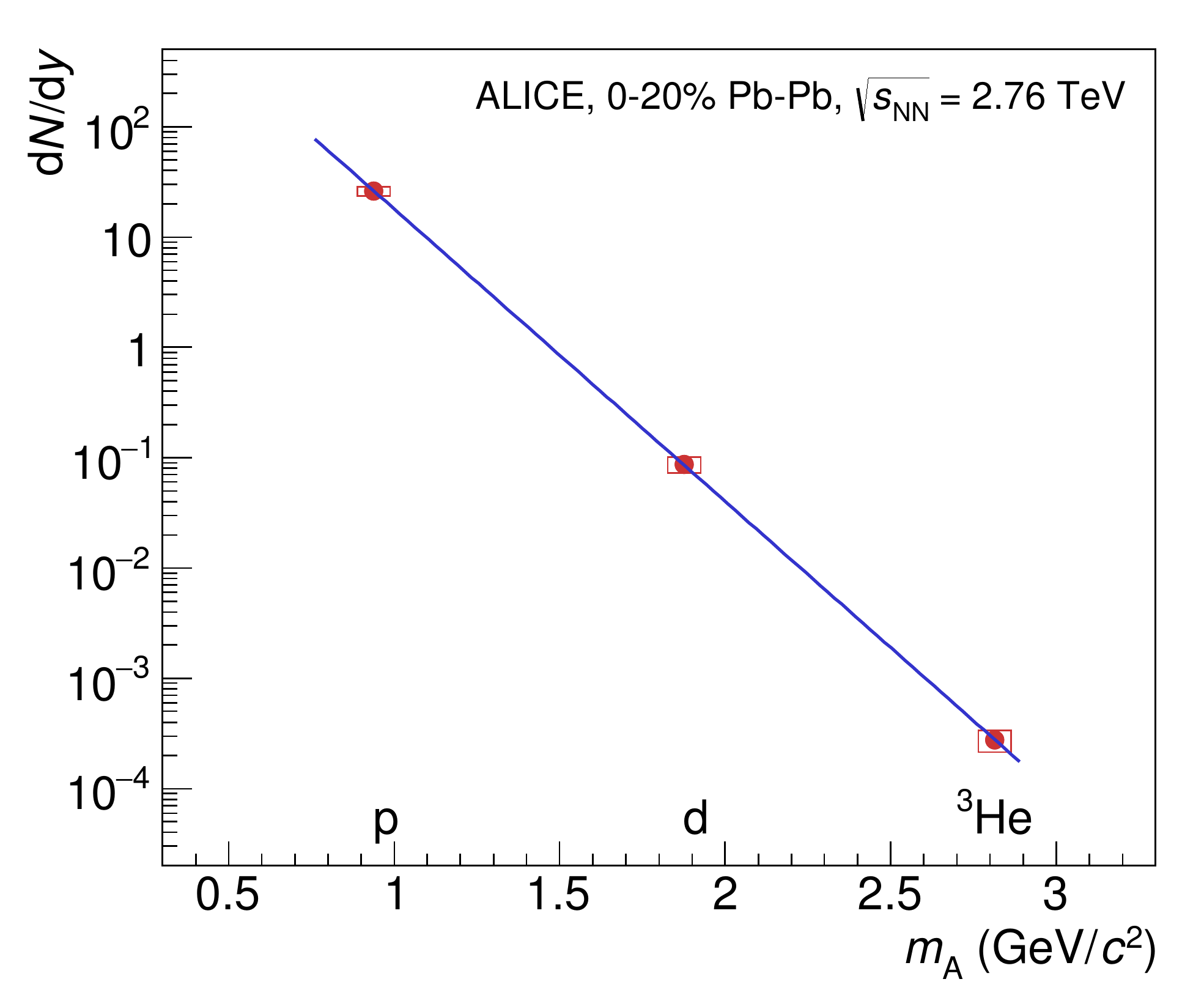}
\caption{(Color online) The production yield d$N$/d$y$ of light nuclei as a
function of the particle mass $m_{{\rm A}}$ measured for 0-20\%
centrality class in Pb--Pb collisions at \snn\ = 2.76 TeV. The
line represents a fit with an exponential function.}
\label{yield-mass}
\end{center}
\end{figure}

Figure~\ref{yield-mass} shows the production yields of  p, d, and
$\rm ^3He$ measured in the centrality interval 0--20\% in Pb--Pb
collisions which follow an exponential decrease with the mass of
the particle.
The penalty factor, namely the reduction
of the yield by adding one nucleon, is 307 $\pm$ 76. Such an exponential decrease
has already been observed at lower incident energies starting from those provided by the
AGS~\cite{Barrette:1994tw,BraunMunzinger:1994iq,Arsenescu:2003eg,Agakishiev:2011ib}, yet with different slopes.

\begin{figure}[h]
\begin{center}
 \includegraphics[width=8.5cm]{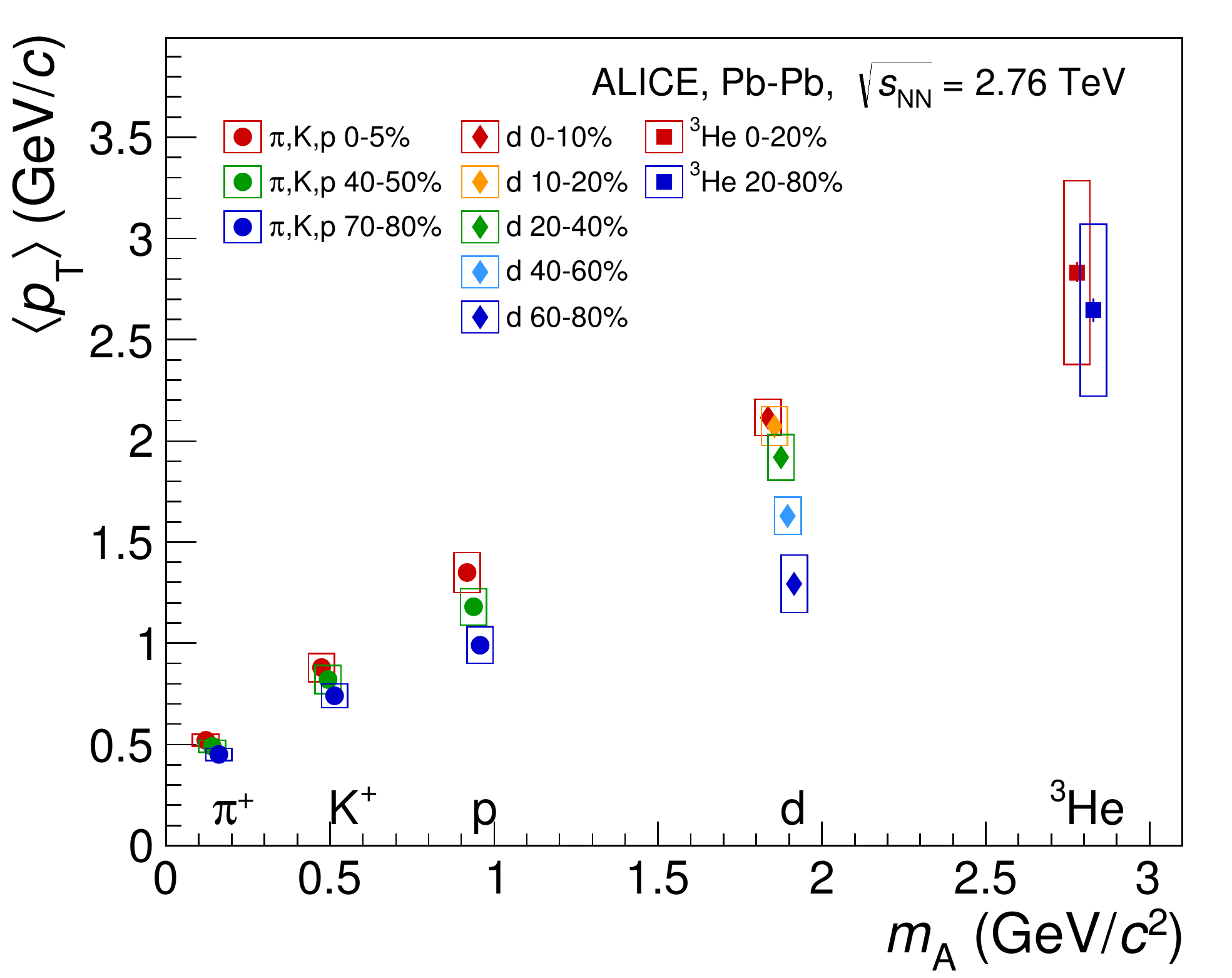}
 \caption{(Color online) Mean transverse momentum $\langle p_{\rm T} \rangle$ as a
   function of particle mass for various centrality classes are shown for Pb--Pb collisions at \snn\ = 2.76 TeV.}
\label{meanpt_he3}
\end{center}
\end{figure}

The mean transverse momentum $\langle p_{\rm T} \rangle$ values
obtained for d and $\rm ^3He$ are compared to those of light
particle species for Pb--Pb collisions at \snn\ = 2.76 TeV
(from~\cite{Abelev:2013vea}) in Fig.~\ref{meanpt_he3}.
The figure shows that the $\langle p_{\rm T} \rangle$ increases
with increasing mass of the particle.
Such a behavior is expected if all the particles are emitted from a
radially expanding
source.

\subsection{\label{secAntiTriton}Observation of (anti-)triton}

\begin{figure}[h]
 \begin{center}
  \includegraphics[width=8.5cm]{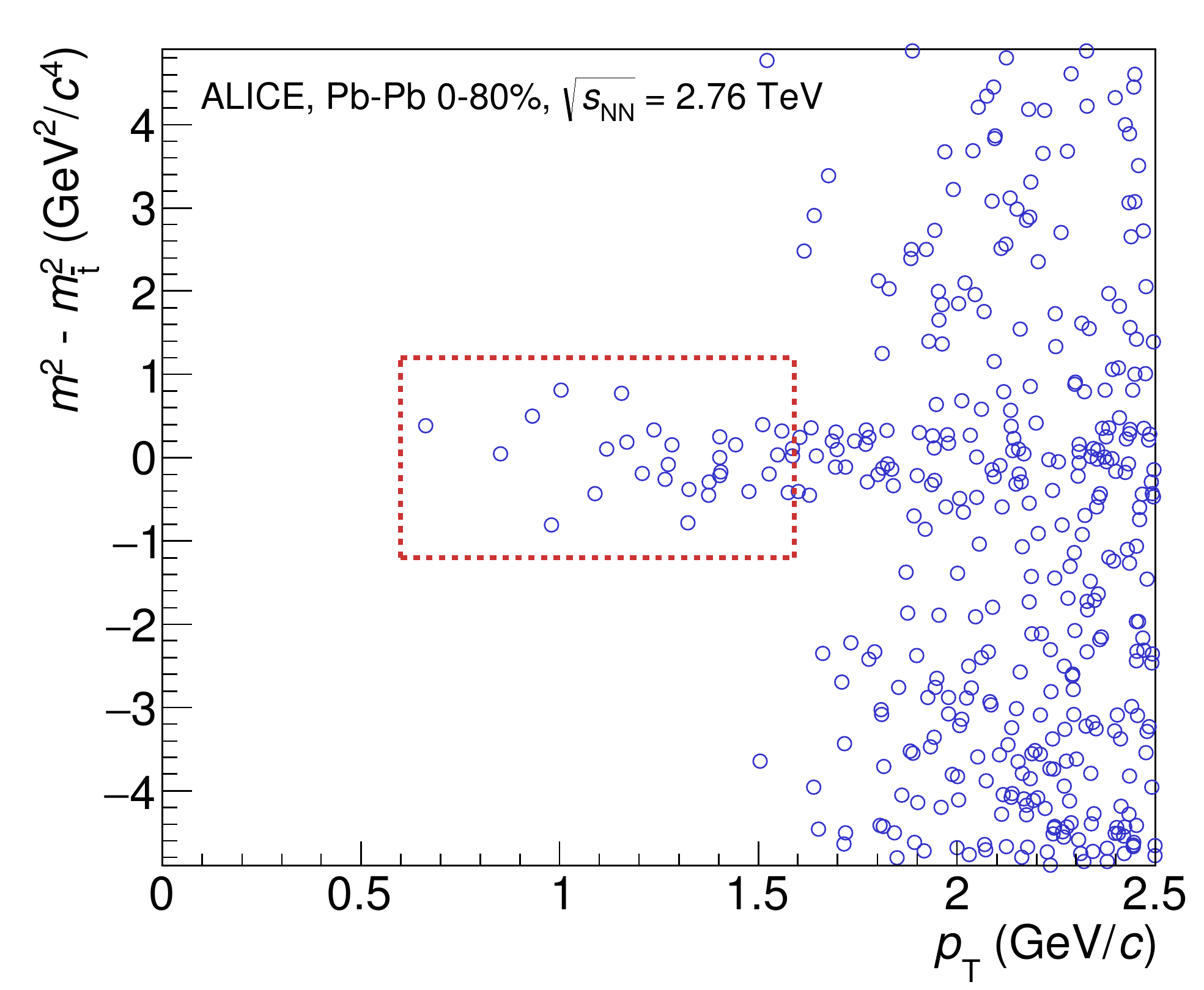}
\caption{(Color online) Scatter plot of $(m^{2} - m^{2}_{\rm \bar{t}})$ measured with the TOF detector versus $p_{\mathrm{T}}$.
Only those tracks are shown which pass the pre-selection done by applying a 3$\sigma$ cut on the TPC d$E$/d$x$. 
The $p_{\mathrm{T}}$-region in which the candidates are identified on a track-by-track basis is shown as red box.}
  \label{fig.:anti-triton}
 \end{center}
\end{figure}

The combined particle identification capability of the TPC and TOF also allows a
track-by-track identification of low momenta (0.6 GeV/$c$ $<$
$p_{\mathrm{T}}$ $<$ 1.6 GeV/$c$) anti-tritons as illustrated in
Fig.~\ref{fig.:anti-triton}. In this momentum region, the background from
mismatched tracks is removed by the TPC particle identification. The contamination 
is estimated based on a side-band study and found to be negligible below $p_{\mathrm{T}}$~$<$~1.6~GeV/$c$, 
but it increases rapidly for higher momenta so that signal and background cannot be distinguished anymore
thus limiting the range available for the measurement.

As can be seen, 31 anti-triton candidates are
observed in the 0-80\% centrality range. These numbers are
consistent with expectations based on an extrapolation of the $\rm
^3He$-spectra to lower momenta taking into account the low
reconstruction efficiency for anti-tritons in this momentum region
(of about 11\% $\pm$ 6\%). An observation of about 10 to 40
anti-tritons is expected based on this estimate, indicating
similar production rates of anti-tritons and
$^{3}{\mathrm{\overline{He}}}$ nuclei. This comparison suffers
from large uncertainties related to the absorption of anti-nuclei
and energy loss in the detector material before the TPC at such 
low momenta. A similar measurement of tritons is unfeasible due
to the large contamination from knock-out nuclei in this momentum
region.

\section{\label{secDis} Discussion}

\subsection{Description of spectra via blast-wave fits}

\begin{figure}[tbp]
 \begin{center}
  \includegraphics[width=9.0cm]{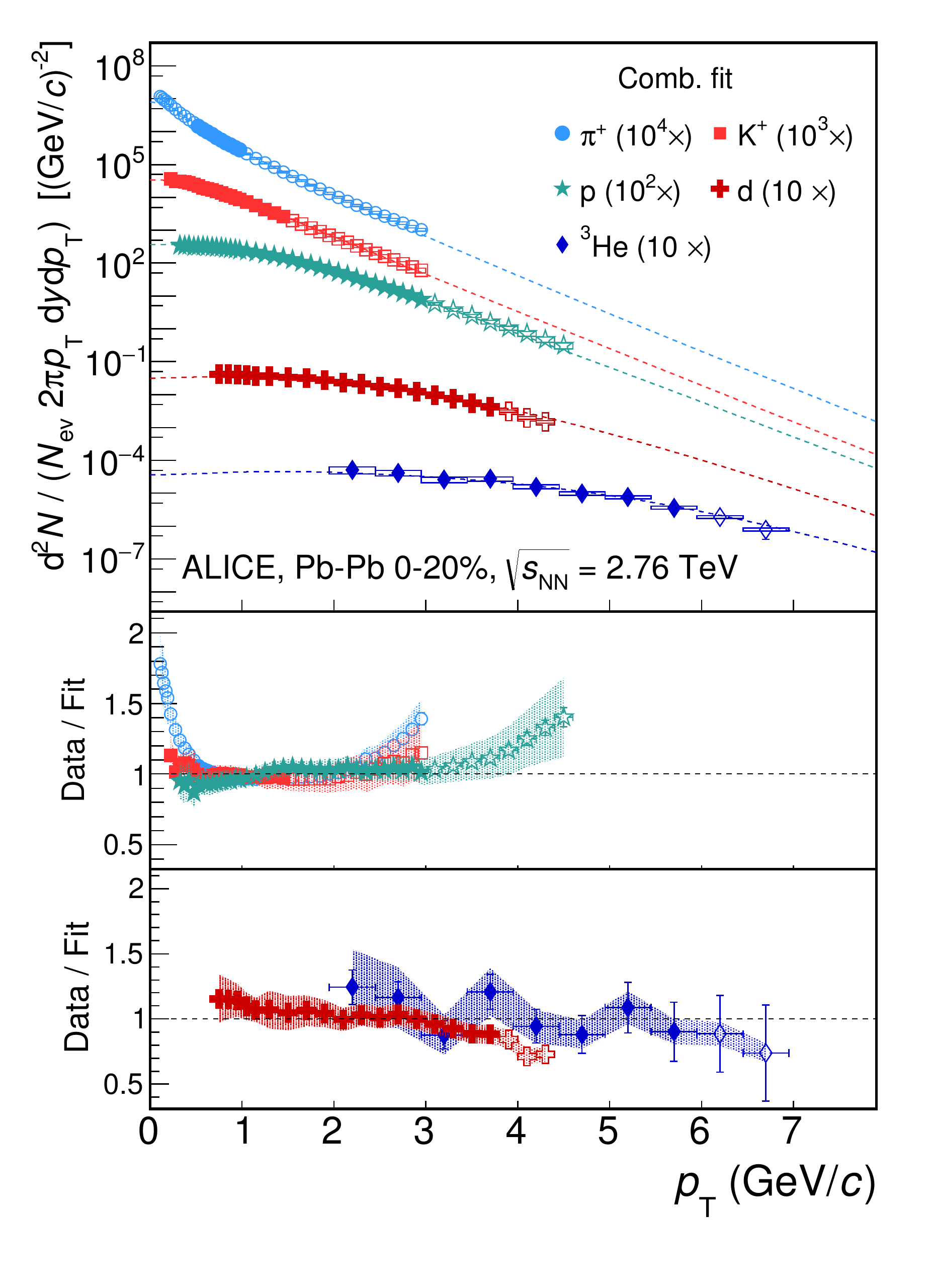}

 \caption{(Color online) Blast-wave fit of $\pi^+$, K$^+$, p, 
  d, and $\rm ^{3}He$ particles for 0--20\% centrality for Pb--Pb collisions at $\sqrt{s_{\rm NN}}$ = 2.76 TeV. 
Solid symbols denote the $p_{\rm T}$ range of the spectra used for
the fits, while the open symbols show the remaining part.
The lower
panels show the deviations of the measured spectra to the BW fits.}
\label{CombinedBW}
 \end{center}
\end{figure}

Combined BW fits provide essential insight into the kinetic freeze-out conditions
and allow quantitative comparisons between different collision systems and between measurements 
at different \snn\ in terms of a hydrodynamic interpretation. In this section, a simultaneous fit to the $\pi$, K, p,
d, and $^{3}$He spectra in the centrality range 0-20\% using in addition data from~\cite{Abelev:2012wca,Abelev:2013vea} 
is discussed. Since the BW model
is not expected to describe eventual hard contributions that may set in at higher
\pt, the fit ranges have been limited.
For the light particles, they are taken as
in~\cite{Abelev:2013vea,Abelev:2012wca} (0.5-1 GeV/$c$, 0.2-1.5
GeV/$c$, 0.3-3 GeV/$c$ for $\pi$, K, and p, respectively).  
However, for 
d and $^{3}$He, 
the spectrum is fitted up to the
\pt~value where the invariant yield reduces to 10\% of the maximum
available value of that spectrum. The exponent $n$ of the velocity profile
is left as a free parameter as in \cite{Abelev:2013vea}. In such an approach,
all particle species are forced to decouple with the same parameters even though
they feature different hadronic cross sections with the medium. This is in particular
relevant for multi-strange particles such as $\Xi$ and $\Omega$ \cite{Adams:2005dq},
which are therefore not included in the fit. 

In Fig.~\ref{CombinedBW} the results of a simultaneous fit to the
five particle
species
are shown. 
The deviations of the spectra from the BW fit are shown in the
lower parts of Fig.~\ref{CombinedBW}. The statistical errors are shown by vertical lines and the systematic
errors are shown as shaded bands. Note that data points marked with open symbols
are not included in the fit.
The hardening of the spectra for central collisions is qualitatively well described 
by the combined BW fit with a collective radial flow velocity $\langle
\beta \rangle$ = 
0.632 $\pm$ 0.01, 
a
kinetic freeze-out temperature of $T_{\rm kin}$ = 
113 $\pm$ 12
MeV, and $n$ = 
0.72 $\pm$ 0.03. 
The $\chi^{2}/{\rm ndf}$ value of
the fit is 0.4. 
A comparison of these parameters to those obtained from a fit to $\pi$, K, and p \cite{Abelev:2013vea} 
($\langle \beta \rangle$~=~0.644~$\pm$~0.020, $T_{\rm kin}$ = 97 $\pm$ 15 MeV, and $n$~=~0.73~$\pm$~0.11) reveals that
the inclusion of nuclei leads to a slightly smaller value for $\langle \beta \rangle$  and a slightly larger value 
for $T_{\rm kin}$. This behavior is mainly driven by the strong anti-correlation of 
$\langle \beta \rangle$ and $T_{\rm kin}$ in the blast-wave model:
the slightly lower value of $\langle \beta \rangle$ leads to a deviation
of the fit from the proton spectrum which is then compensated by a higher
$T_{\rm kin}$. 


\subsection{Comparison to thermal models}

\begin{figure}
 \begin{center}
 \includegraphics[width=8.5cm]{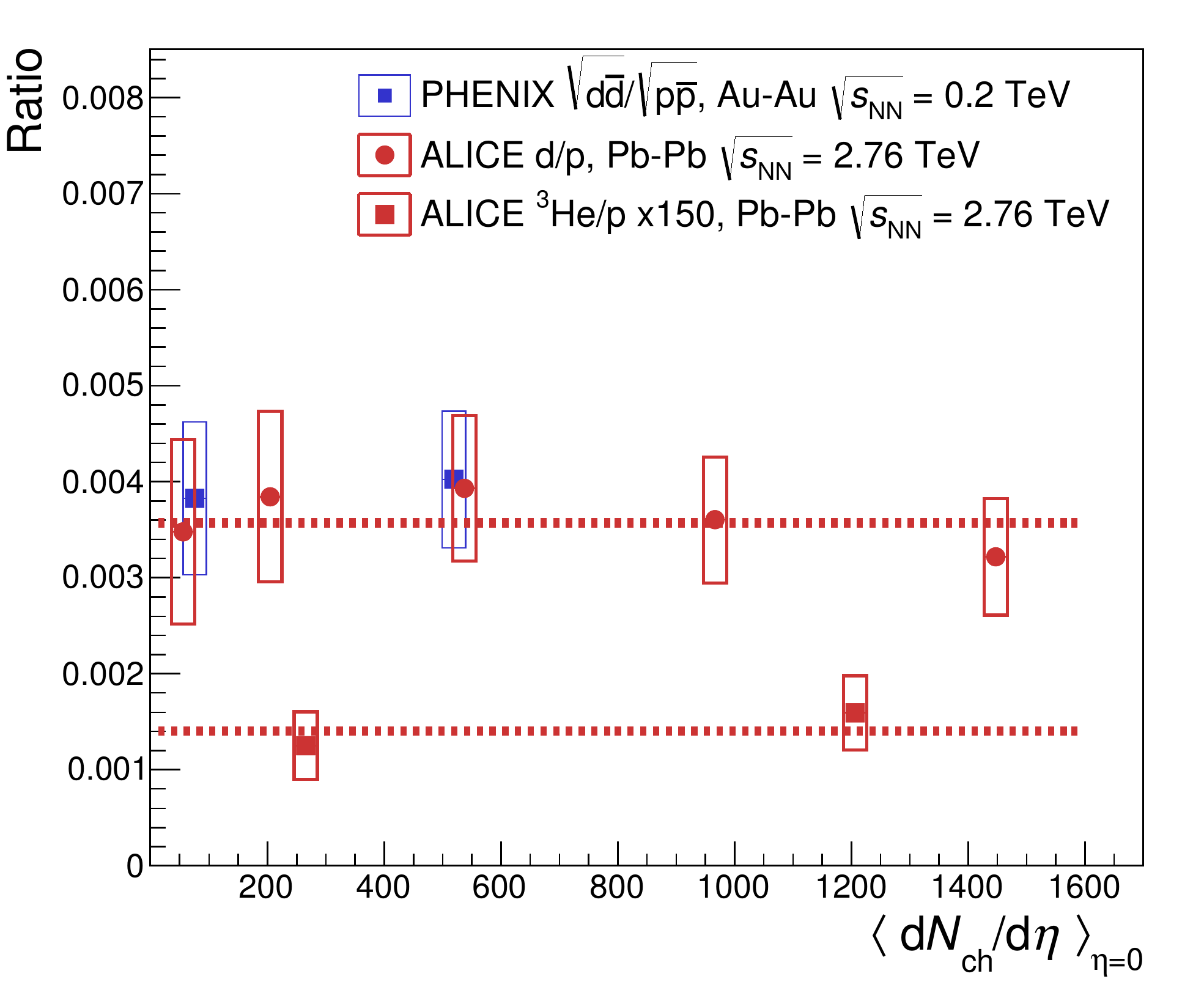}
     \caption{(Color online) d/p and $^{3}$He/p ratio in heavy-ion collisions as a function of event multiplicity.
  Within the uncertainties no significant variation with multiplicity is observed.
  The d/p and \dbar/\pbar\ results from the PHENIX Collaboration~\cite{Adler:2004uy,Adler:2003cb}
  are averaged as explained in the text. The lines represent fits with
  a constant 
  to the ALICE data points.}
   \label{figParticleRatios}
 \end{center}
\end{figure}

Figure~\ref{figParticleRatios} shows the d/p and the
$^{3}$He/p ratios as a function of the average charged particle multiplicity per event. 
The proton yields are taken from~\cite{Abelev:2012wca,Abelev:2013vea}.
The observed values of about $3.6\times 10^{-3}$ for the d/p ratio
and about $9.3 \times 10^{-6}$ for the $^{3}$He/p ratio are in
agreement with expectations from the thermal-statistical 
models~\cite{Cleymans:2011pe,Andronic:2010qu}. Similar values for d/p
ratios are
also observed by the PHENIX experiment for Au-Au
collisions~\cite{Adler:2004uy,Adler:2003cb}. Since at RHIC energies
significant differences between nucleus and anti-nucleus production are present,
for this plot the geometrical mean is used which in a thermal
concept cancels the influence of the baryon chemical potential
($\mu_{\rm B}$) \footnotemark[1].\footnotetext[1]{In a thermal model, the yield $n_{B}$ of a baryon with energy $E$ in a medium
of temperature $T$ is proportional to $\exp(-{E - \mu_{B} \over T})$   while the
yield of an anti-baryon $n_{\overline{B}}$ is proportional $\exp(-{E + \mu_{B} \over T})$.
The geometric mean $\sqrt{n_{B}n_{\overline{B}}}$ leads to a cancellation of the $\mu_{B}$.}
Within the achieved experimental precision, no
dependence of these particle ratios on the event multiplicity is
observed at RHIC and LHC energies.
Also the \pbar/p and the p/$\pi$ ratios hardly vary with
centrality~\cite{Abelev:2008ez,Abelev:2013vea} showing that
$T_{\rm chem}$ and $\mu_{\rm B}$ do not vary with centrality in
high energy collisions.
In a coalescence approach, the centrality independence disfavors
implementations in which the nuclei production is proportional to
the absolute proton multiplicity~\cite{PhysRevC.88.034908} rather
than the particle density.

\begin{figure}
 \begin{center}
   \includegraphics[width=8.5cm]{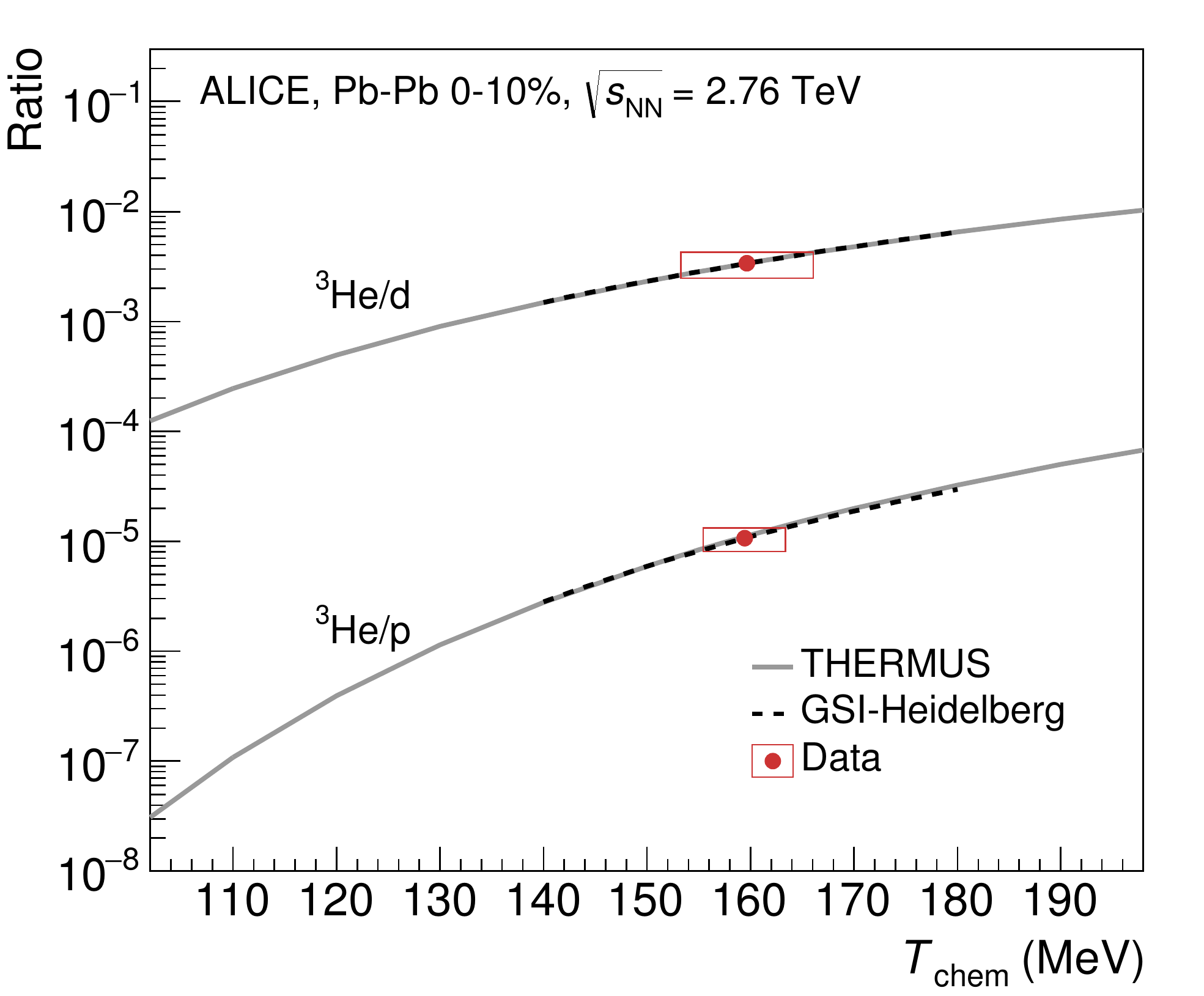}
  \caption{(Color online) Particle ratios of nuclei 
  as measured in 0--10\% most
  central Pb--Pb collisions compared to the THERMUS~\cite{Wheaton:2004qb} model (solid lines)
  and the GSI-Heidelberg model~\cite{Andronic:2010qu} (dashed lines)
  as a function of the chemical
  freeze-out temperature $T_{\rm chem}$. The \he\ yield is scaled to 0--10\%. Horizontal error bars indicate
  the temperature range obtained by a projection of the total error of the ratio on the temperature axis.}
  \label{thermus_ratio}
 \end{center}
\end{figure}

The comparison with thermal models is shown in more detail in
Fig.~\ref{thermus_ratio} for the 0--10\% centrality class.
These calculations have been performed using the grand-canonical
formulation of both THERMUS~\cite{Wheaton:2004qb} and the
GSI-Heidelberg model~\cite{Andronic:2010qu}. This approach is appropriate 
for the ratios shown here, as no strange quarks are involved.  Details
can be found in~\cite{Cleymans:2011pe,Andronic:2010qu}. These
ratios are monotonically increasing with $T_{\rm chem}$ reflecting
the dependence with $\exp(-\Delta m/ T_{\rm chem})$ where $\Delta
m$ corresponds to the mass difference of the particles under
study.

The measured ratios of \he/p and \he/d are in agreement with a
chemical freeze-out temperature in the range of 150~MeV to 165~MeV.
No significant differences are observed between the THERMUS and
GSI-Heidelberg model with respect to the production of light \mbox{(anti-)}nuclei.
A fit to p, d, and \he\ only gives
$T_{\rm chem}$ = $156\pm4$ MeV with a $\chi^{2}/{\rm ndf}$ of 0.4.
This value can be
compared to a fit including all measured light flavor hadrons
which yields a temperature of about 156~MeV~\cite{Stachel:2013zma}.

At these temperatures, the weakly-bound deuteron and \he\ can
hardly survive. These nuclei might break up and might be
regenerated. However, if this complex process of break-up and regeneration 
is governed by an overall isentropic expansion, the
particle ratios are preserved~\cite{Siemens:1979dz}. 
Eventually, the yields of particles including weakly bound nuclei
are therefore described in the thermal-statistical model. Other properties,
e.g.~spectral shapes and elliptic flow, exhibit the influence of the
interactions during the hadronic phase.

The d/p ratio obtained in pp collisions is lower by a factor of
2.2 than in Pb--Pb collisions. Assuming thermal production not only in Pb--Pb, 
but also in pp collisions, this could indicate a lower freeze-out temperature in pp collisions.
However, the p/$\pi$ ratio does not show significant
differences between pp and Pb--Pb collisions. 
Effects related to canonical suppression of strange particles can also
be excluded because these ratios do not involve any strange quarks.
Therefore, this
observation must find another explanation within the framework of thermal models
or non-thermal production mechanisms need to be considered in small systems.
Further work in the theoretical models is needed for a better
understanding of this effect.

\subsection{Comparison with the coalescence model}
\label{subsec.:Coalescence}

Light nuclei have nucleons as constituents and are thus likely formed 
via coalescence of protons and neutrons which are near
in space and have similar velocities. In this production
mechanism, the spectral distribution of the composite nuclei is
related to the one of the primordial nucleons via

\begin{equation}
E_i \, \frac{{\rm d^3} N_i}{{\rm d} p_i^3} =  B_A \, \left(
E_{\rm p} \, \frac{{\rm d^3} N_{\rm p}}{{\rm d} p_{\rm p}^3}
\right)^A
\label{coal}
\end{equation}
\noindent assuming that protons and neutrons have the same momentum
distribution. $B_A$ is the coalescence parameter for nuclei $i$
with mass number $A$ and a momentum of $p_i = A \, p_{\rm p}$.





\begin{figure}[h]
\includegraphics[width=8.0cm]{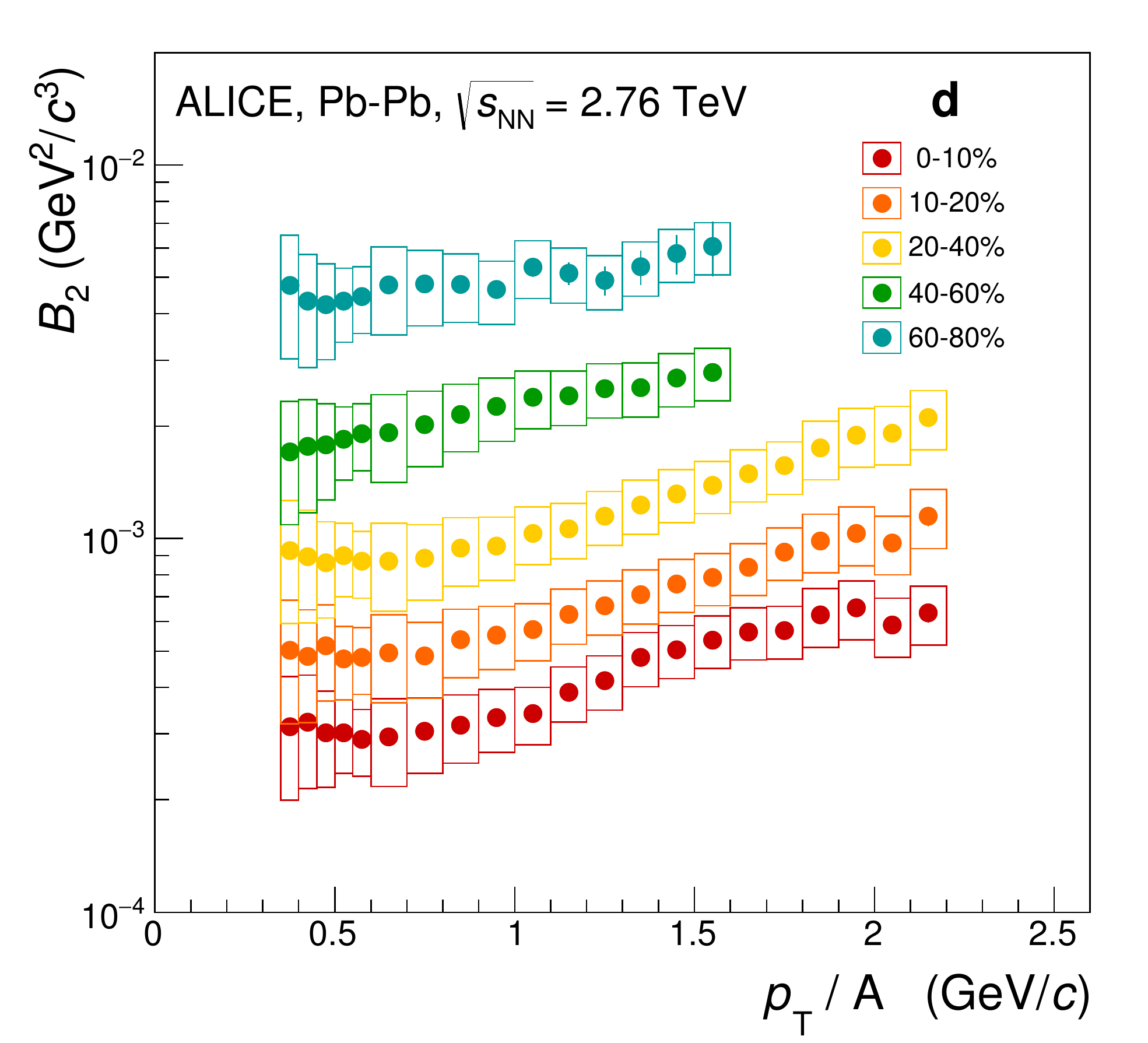}
 \includegraphics[width=8.0cm]{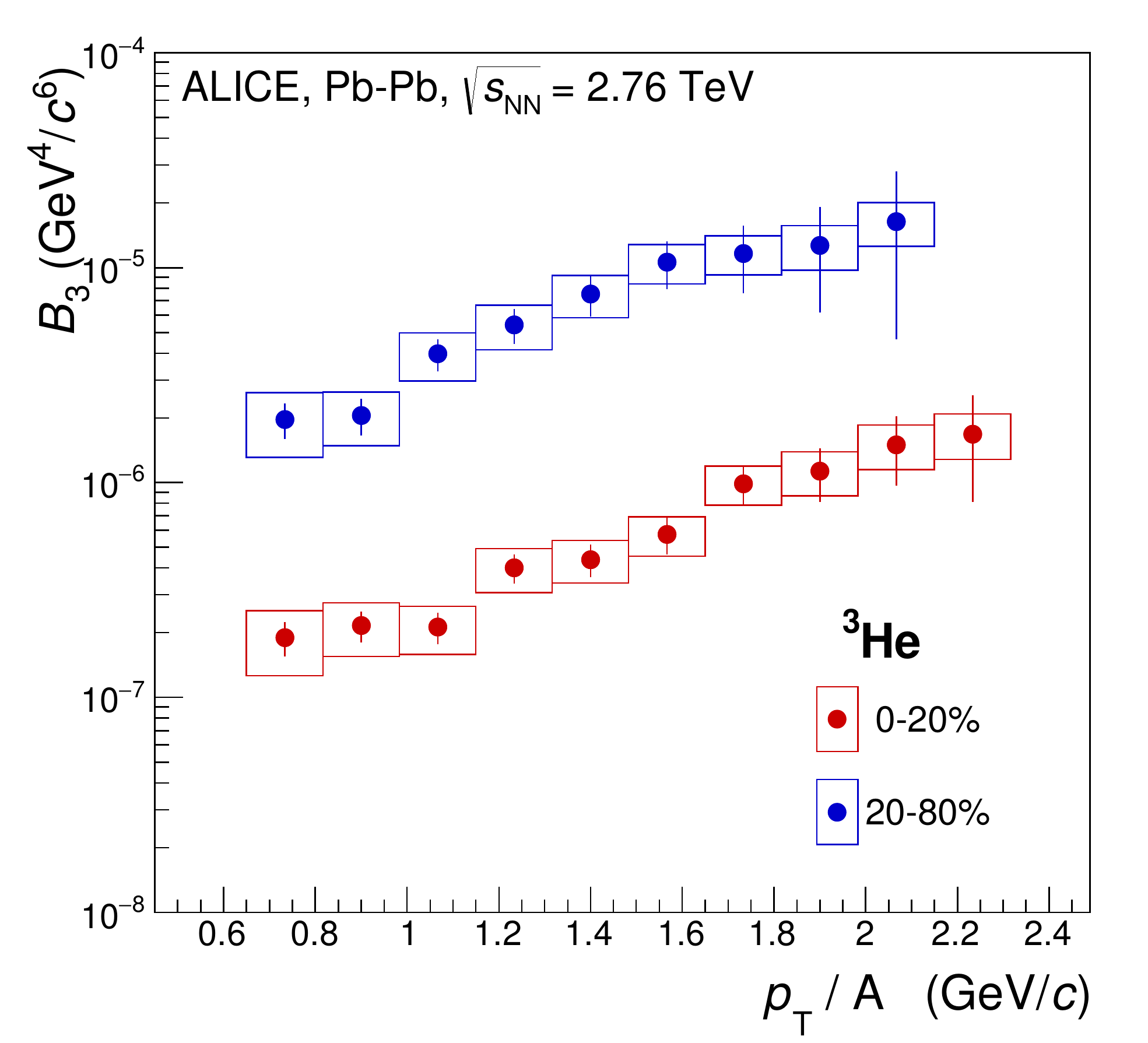}
\caption{(Color online) The coalescence parameters $B_2$ (left) and $B_3$ (right)
as a function of the transverse momentum per nucleon for various
centrality classes in
Pb--Pb collisions at \snn\ = 2.76 TeV.} 
\label{B2B3}
\end{figure}

Figure~\ref{B2B3} shows the obtained $B_{2}$ values for deuterons
(left panel) and $B_{3}$ values for \he\ (right panel) in several
centrality bins for Pb--Pb collisions. 
The results are plotted versus the transverse momentum per nucleon.
A clear decrease of $B_{2}$ and $B_{3}$ with increasing centrality is observed. In the coalescence
picture, this behavior is explained by an increase in the
source volume $V_{\rm eff}$: the larger the distance between the
protons and neutrons which are created in the collision, the less
likely is that they coalesce. Alternatively, it can be understood
on the basis of the approximately constant d/p and \he/p-ratios as an increase
of the overall proton multiplicity independent of the geometry of the
collision. The argument can be best illustrated by assuming a
constant value of $B_{2}$ and integrating Eq.~(\ref{coal}) over
\pt. The value of $B_{2}$ can then be calculated for a given ratio
d/p and a given spectral shape $f(p_{\rm T})$ (with
$\int_{0}^{\infty}f(p_{\rm T})\, {\rm d} p_{\rm T} = 1$) of the
proton spectrum as

\begin{equation}
B_{2} = {\pi \over 2} \cdot { {{\rm d}N_{\rm d} \over {\rm d}y}
\over {({ {\rm d}N_{\rm p} \over {\rm d}y})^{2}}} \cdot { 1 \over
\int_{0}^{\infty} {f^{2}(p_{\rm T}) \over p_{\rm T}} {\rm d}
p_{\rm T}} \ ,
\label{coal2}
\end{equation}

\noindent where for a constant ratio of the deuteron ${{\rm
d}N_{\rm d} / {\rm d}y}$ to proton ${{\rm d}N_{\rm p} / {\rm d}y}$
yield, it is found that $B_{2} \propto 1/({{\rm d}N_{\rm p} /{\rm
d}y})$. As can be seen in Fig.~\ref{B2B3}, the coalescence parameter also
develops an increasing trend with transverse momentum for
central collisions in contrast to expectations of the most simple coalescence
models. The significance of this increase is further substantiated by the
fact that the systematic errors between \pt-bins are to a large extent correlated.
It can be qualitatively explained by position-momentum correlations which are
caused by a radially expanding source \cite{Polleri:1997bp}.
For quantitative comparisons, better theoretical calculations are needed which couple
a coalescence model to a full space-time hydrodynamic description of the fireball. Also in the discussion 
of the variation of the $B_{2}$ parameter as a function of collision energy, 
its strong dependence on centrality and \pt\ must be taken into account. It
is observed that $B_{2}$ at a fixed momentum (\pt~=~1.3~GeV/$c$)
for central collisions (0-20\%) decreases rapidly from AGS
energies to top SPS energy and then remains about the same up to
RHIC~\cite{Adler:2004uy}. Our value of approximately $4\times
10^{-4}$ GeV$^{2}$/$c^{3}$ is only slightly lower than the
measurement at RHIC ($\approx 6\times10^{-4}$
GeV$^{2}$/$c^{3}$).

\section{\label{secConclusion} Conclusion}

In summary, the spectral distributions of deuterons in pp at \s\ =
7 TeV and of deuterons and \he\ in Pb--Pb collisions at \snn\ =
2.76 TeV have been presented. In Pb--Pb collisions, the yields are decreasing 
by a factor of 307 $\pm$ 76 for each additional nucleon, the mean \pt\ rises with mass
and 
the combined blast-wave fit to $\pi$, K, p, d, and $^{3}$He
gives a reasonable fit with $\langle \beta \rangle$ = 0.63 and $T_{\rm kin}$ around
115 MeV suggesting 
that the kinetic freeze-out conditions for nuclei are identical to those of the other light flavour hadrons.
For anti-tritons, a track-by-track
identification has been applied in the momentum range 0.6 GeV/$c$ $<$ $p_{\mathrm{T}}$ $<$ 1.6 GeV/$c$
and the observation of 31 anti-tritons in Pb--Pb collisions
at \snn\ = 2.76 TeV in the 0-80\% centrality class is reported in this paper.

An important question is whether the nuclei produced in heavy-ion collisions
are created at the chemical freeze-out or at a later stage via coalescence.
One of the key observations is the fact that the d/p and \he/p ratios
are constant as a function of $\langle$d$N_{\rm ch}$/d$\eta$$\rangle_{\eta=0}$. 
Such a behavior
is expected from a thermal-statistical interpretation, as it is
found that $T_{\rm chem}$ and $\mu_{\rm B}$ do not vary with
centrality in high energy collisions. 
Furthermore, a common freeze-out temperature of around 156 MeV
for light \mbox{(anti-)}nuclei and all other measured light flavor hadrons is obtained in a thermal-statistical model.
Light \mbox{(anti-)}nuclei in
Pb--Pb collisions therefore show the identical behavior as non-composite light flavor
hadrons which are governed by a common chemical freeze-out and a subsequent hydrodynamic
expansion.

The extracted coalescence parameters $B_2$ and $B_3$ exhibit a significant decrease with collision centrality and an increase with transverse momentum which cannot be explained by coalescence models in their simplest form. On the other hand, taking into account the larger source volume for more central collisions and the radial expansion of the emitting system, 
the production of light \mbox{(anti-)}nuclei in Pb--Pb collisions may still be compatible with the expectations from a coalescence space-time description.

The measurements of nuclei at LHC energies are shown to
follow trends observed from lower incident energies.
Extrapolations and model predictions based on the thermal-statistical or
coalescence approach are, therefore, a solid
ground for further studies, e.g.~of hyper-nuclei and exotica.

\newenvironment{acknowledgement}{\relax}{\relax}
\begin{acknowledgement}
\section*{Acknowledgements}

The ALICE Collaboration would like to thank all its engineers and technicians for their invaluable contributions to the construction of the experiment and the CERN accelerator teams for the outstanding performance of the LHC complex.
The ALICE Collaboration gratefully acknowledges the resources and support provided by all Grid centres and the Worldwide LHC Computing Grid (WLCG) collaboration.
The ALICE Collaboration acknowledges the following funding agencies for their support in building and
running the ALICE detector:
State Committee of Science,  World Federation of Scientists (WFS)
and Swiss Fonds Kidagan, Armenia,
Conselho Nacional de Desenvolvimento Cient\'{\i}fico e Tecnol\'{o}gico (CNPq), Financiadora de Estudos e Projetos (FINEP),
Funda\c{c}\~{a}o de Amparo \`{a} Pesquisa do Estado de S\~{a}o Paulo (FAPESP);
National Natural Science Foundation of China (NSFC), the Chinese Ministry of Education (CMOE)
and the Ministry of Science and Technology of China (MSTC);
Ministry of Education and Youth of the Czech Republic;
Danish Natural Science Research Council, the Carlsberg Foundation and the Danish National Research Foundation;
The European Research Council under the European Community's Seventh Framework Programme;
Helsinki Institute of Physics and the Academy of Finland;
French CNRS-IN2P3, the `Region Pays de Loire', `Region Alsace', `Region Auvergne' and CEA, France;
German Bundesministerium fur Bildung, Wissenschaft, Forschung und Technologie (BMBF) and the Helmholtz Association;
General Secretariat for Research and Technology, Ministry of
Development, Greece;
Hungarian Orszagos Tudomanyos Kutatasi Alappgrammok (OTKA) and National Office for Research and Technology (NKTH);
Department of Atomic Energy and Department of Science and Technology of the Government of India;
Istituto Nazionale di Fisica Nucleare (INFN) and Centro Fermi -
Museo Storico della Fisica e Centro Studi e Ricerche "Enrico
Fermi", Italy;
MEXT Grant-in-Aid for Specially Promoted Research, Ja\-pan;
Joint Institute for Nuclear Research, Dubna;
National Research Foundation of Korea (NRF);
Consejo Nacional de Cienca y Tecnologia (CONACYT), Direccion General de Asuntos del Personal Academico(DGAPA), M\'{e}xico, Amerique Latine Formation academique - European Commission~(ALFA-EC) and the EPLANET Program~(European Particle Physics Latin American Network);
Stichting voor Fundamenteel Onderzoek der Materie (FOM) and the Nederlandse Organisatie voor Wetenschappelijk Onderzoek (NWO), Netherlands;
Research Council of Norway (NFR);
National Science Centre, Poland;
Ministry of National Education/Institute for Atomic Physics and National Council of Scientific Research in Higher Education~(CNCSI-UEFISCDI), Romania;
Ministry of Education and Science of Russian Federation, Russian
Academy of Sciences, Russian Federal Agency of Atomic Energy,
Russian Federal Agency for Science and Innovations and The Russian
Foundation for Basic Research;
Ministry of Education of Slovakia;
Department of Science and Technology, South Africa;
Centro de Investigaciones Energeticas, Medioambientales y Tecnologicas (CIEMAT), E-Infrastructure shared between Europe and Latin America (EELA), Ministerio de Econom\'{i}a y Competitividad (MINECO) of Spain, Xunta de Galicia (Conseller\'{\i}a de Educaci\'{o}n),
Centro de Aplicaciones Tecnológicas y Desarrollo Nuclear (CEA\-DEN), Cubaenerg\'{\i}a, Cuba, and IAEA (International Atomic Energy Agency);
Swedish Research Council (VR) and Knut $\&$ Alice Wallenberg
Foundation (KAW);
Ukraine Ministry of Education and Science;
United Kingdom Science and Technology Facilities Council (STFC);
The United States Department of Energy, the United States National
Science Foundation, the State of Texas, and the State of Ohio;
Ministry of Science, Education and Sports of Croatia and  Unity through Knowledge Fund, Croatia.
Council of Scientific and Industrial Research (CSIR), New Delhi, India

\end{acknowledgement}

\bibliographystyle{utphys} 	
\bibliography{antinuclei}

\newpage
\appendix

\newpage
\section{The ALICE Collaboration}
\label{app:collab}



\begingroup
\small
\begin{flushleft}
J.~Adam\Irefn{org40}\And
D.~Adamov\'{a}\Irefn{org83}\And
M.M.~Aggarwal\Irefn{org87}\And
G.~Aglieri Rinella\Irefn{org36}\And
M.~Agnello\Irefn{org111}\And
N.~Agrawal\Irefn{org48}\And
Z.~Ahammed\Irefn{org131}\And
I.~Ahmed\Irefn{org16}\And
S.U.~Ahn\Irefn{org68}\And
I.~Aimo\Irefn{org94}\textsuperscript{,}\Irefn{org111}
% ...
\renewcommand\labelenumi{\textsuperscript{\theenumi}~}

\section*{Affiliation notes}
\renewcommand\theenumi{\roman{enumi}}
\begin{Authlist}
\item \Adef{0}Deceased
\item \Adef{idp3781232}{Also at: M.V. Lomonosov Moscow State University, D.V. Skobeltsyn Institute of Nuclear, Physics, Moscow, Russia}
\item \Adef{idp5903136}{Also at: University of Kansas, Lawrence, Kansas, United States}
\end{Authlist}

\section*{Collaboration Institutes}
\renewcommand\theenumi{\arabic{enumi}~}
\begin{Authlist}

\item \Idef{org1}A.I. Alikhanyan National Science Laboratory (Yerevan Physics Institute) Foundation, Yerevan, Armenia
\item \Idef{org2}Benem\'{e}rita Universidad Aut\'{o}noma de Puebla, Puebla, Mexico
\item \Idef{org3}Bogolyubov Institute for Theoretical Physics, Kiev, Ukraine
\item \Idef{org4}Bose Institute, Department of Physics and Centre for Astroparticle Physics and Space Science (CAPSS), Kolkata, India
\item \Idef{org5}Budker Institute for Nuclear Physics, Novosibirsk, Russia
\item \Idef{org6}California Polytechnic State University, San Luis Obispo, California, United States
\item \Idef{org7}Central China Normal University, Wuhan, China
\item \Idef{org8}Centre de Calcul de l'IN2P3, Villeurbanne, France
\item \Idef{org9}Centro de Aplicaciones Tecnol\'{o}gicas y Desarrollo Nuclear (CEADEN), Havana, Cuba
\item \Idef{org10}Centro de Investigaciones Energ\'{e}ticas Medioambientales y Tecnol\'{o}gicas (CIEMAT), Madrid, Spain
\item \Idef{org11}Centro de Investigaci\'{o}n y de Estudios Avanzados (CINVESTAV), Mexico City and M\'{e}rida, Mexico
\item \Idef{org12}Centro Fermi - Museo Storico della Fisica e Centro Studi e Ricerche ``Enrico Fermi'', Rome, Italy
\item \Idef{org13}Chicago State University, Chicago, Illinois, USA
\item \Idef{org14}China Institute of Atomic Energy, Beijing, China
\item \Idef{org15}Commissariat \`{a} l'Energie Atomique, IRFU, Saclay, France
\item \Idef{org16}COMSATS Institute of Information Technology (CIIT), Islamabad, Pakistan
\item \Idef{org17}Departamento de F\'{\i}sica de Part\'{\i}culas and IGFAE, Universidad de Santiago de Compostela, Santiago de Compostela, Spain
\item \Idef{org18}Department of Physics and Technology, University of Bergen, Bergen, Norway
\item \Idef{org19}Department of Physics, Aligarh Muslim University, Aligarh, India
\item \Idef{org20}Department of Physics, Ohio State University, Columbus, Ohio, United States
\item \Idef{org21}Department of Physics, Sejong University, Seoul, South Korea
\item \Idef{org22}Department of Physics, University of Oslo, Oslo, Norway
\item \Idef{org23}Dipartimento di Elettrotecnica ed Elettronica del Politecnico, Bari, Italy
\item \Idef{org24}Dipartimento di Fisica dell'Universit\`{a} 'La Sapienza' and Sezione INFN Rome, Italy
\item \Idef{org25}Dipartimento di Fisica dell'Universit\`{a} and Sezione INFN, Cagliari, Italy
\item \Idef{org26}Dipartimento di Fisica dell'Universit\`{a} and Sezione INFN, Trieste, Italy
\item \Idef{org27}Dipartimento di Fisica dell'Universit\`{a} and Sezione INFN, Turin, Italy
\item \Idef{org28}Dipartimento di Fisica e Astronomia dell'Universit\`{a} and Sezione INFN, Bologna, Italy
\item \Idef{org29}Dipartimento di Fisica e Astronomia dell'Universit\`{a} and Sezione INFN, Catania, Italy
\item \Idef{org30}Dipartimento di Fisica e Astronomia dell'Universit\`{a} and Sezione INFN, Padova, Italy
\item \Idef{org31}Dipartimento di Fisica `E.R.~Caianiello' dell'Universit\`{a} and Gruppo Collegato INFN, Salerno, Italy
\item \Idef{org32}Dipartimento di Scienze e Innovazione Tecnologica dell'Universit\`{a} del  Piemonte Orientale and Gruppo Collegato INFN, Alessandria, Italy
\item \Idef{org33}Dipartimento Interateneo di Fisica `M.~Merlin' and Sezione INFN, Bari, Italy
\item \Idef{org34}Division of Experimental High Energy Physics, University of Lund, Lund, Sweden
\item \Idef{org35}Eberhard Karls Universit\"{a}t T\"{u}bingen, T\"{u}bingen, Germany
\item \Idef{org36}European Organization for Nuclear Research (CERN), Geneva, Switzerland
\item \Idef{org37}Excellence Cluster Universe, Technische Universit\"{a}t M\"{u}nchen, Munich, Germany
\item \Idef{org38}Faculty of Engineering, Bergen University College, Bergen, Norway
\item \Idef{org39}Faculty of Mathematics, Physics and Informatics, Comenius University, Bratislava, Slovakia
\item \Idef{org40}Faculty of Nuclear Sciences and Physical Engineering, Czech Technical University in Prague, Prague, Czech Republic
\item \Idef{org41}Faculty of Science, P.J.~\v{S}af\'{a}rik University, Ko\v{s}ice, Slovakia
\item \Idef{org42}Faculty of Technology, Buskerud and Vestfold University College, Vestfold, Norway
\item \Idef{org43}Frankfurt Institute for Advanced Studies, Johann Wolfgang Goethe-Universit\"{a}t Frankfurt, Frankfurt, Germany
\item \Idef{org44}Gangneung-Wonju National University, Gangneung, South Korea
\item \Idef{org45}Gauhati University, Department of Physics, Guwahati, India
\item \Idef{org46}Helsinki Institute of Physics (HIP), Helsinki, Finland
\item \Idef{org47}Hiroshima University, Hiroshima, Japan
\item \Idef{org48}Indian Institute of Technology Bombay (IIT), Mumbai, India
\item \Idef{org49}Indian Institute of Technology Indore, Indore (IITI), India
\item \Idef{org50}Inha University, Incheon, South Korea
\item \Idef{org51}Institut de Physique Nucl\'eaire d'Orsay (IPNO), Universit\'e Paris-Sud, CNRS-IN2P3, Orsay, France
\item \Idef{org52}Institut f\"{u}r Informatik, Johann Wolfgang Goethe-Universit\"{a}t Frankfurt, Frankfurt, Germany
\item \Idef{org53}Institut f\"{u}r Kernphysik, Johann Wolfgang Goethe-Universit\"{a}t Frankfurt, Frankfurt, Germany
\item \Idef{org54}Institut f\"{u}r Kernphysik, Westf\"{a}lische Wilhelms-Universit\"{a}t M\"{u}nster, M\"{u}nster, Germany
\item \Idef{org55}Institut Pluridisciplinaire Hubert Curien (IPHC), Universit\'{e} de Strasbourg, CNRS-IN2P3, Strasbourg, France
\item \Idef{org56}Institute for Nuclear Research, Academy of Sciences, Moscow, Russia
\item \Idef{org57}Institute for Subatomic Physics of Utrecht University, Utrecht, Netherlands
\item \Idef{org58}Institute for Theoretical and Experimental Physics, Moscow, Russia
\item \Idef{org59}Institute of Experimental Physics, Slovak Academy of Sciences, Ko\v{s}ice, Slovakia
\item \Idef{org60}Institute of Physics, Academy of Sciences of the Czech Republic, Prague, Czech Republic
\item \Idef{org61}Institute of Physics, Bhubaneswar, India
\item \Idef{org62}Institute of Space Science (ISS), Bucharest, Romania
\item \Idef{org63}Instituto de Ciencias Nucleares, Universidad Nacional Aut\'{o}noma de M\'{e}xico, Mexico City, Mexico
\item \Idef{org64}Instituto de F\'{\i}sica, Universidad Nacional Aut\'{o}noma de M\'{e}xico, Mexico City, Mexico
\item \Idef{org65}iThemba LABS, National Research Foundation, Somerset West, South Africa
\item \Idef{org66}Joint Institute for Nuclear Research (JINR), Dubna, Russia
\item \Idef{org67}Konkuk University, Seoul, South Korea
\item \Idef{org68}Korea Institute of Science and Technology Information, Daejeon, South Korea
\item \Idef{org69}KTO Karatay University, Konya, Turkey
\item \Idef{org70}Laboratoire de Physique Corpusculaire (LPC), Clermont Universit\'{e}, Universit\'{e} Blaise Pascal, CNRS--IN2P3, Clermont-Ferrand, France
\item \Idef{org71}Laboratoire de Physique Subatomique et de Cosmologie, Universit\'{e} Grenoble-Alpes, CNRS-IN2P3, Grenoble, France
\item \Idef{org72}Laboratori Nazionali di Frascati, INFN, Frascati, Italy
\item \Idef{org73}Laboratori Nazionali di Legnaro, INFN, Legnaro, Italy
\item \Idef{org74}Lawrence Berkeley National Laboratory, Berkeley, California, United States
\item \Idef{org75}Lawrence Livermore National Laboratory, Livermore, California, United States
\item \Idef{org76}Moscow Engineering Physics Institute, Moscow, Russia
\item \Idef{org77}National Centre for Nuclear Studies, Warsaw, Poland
\item \Idef{org78}National Institute for Physics and Nuclear Engineering, Bucharest, Romania
\item \Idef{org79}National Institute of Science Education and Research, Bhubaneswar, India
\item \Idef{org80}Niels Bohr Institute, University of Copenhagen, Copenhagen, Denmark
\item \Idef{org81}Nikhef, Nationaal instituut voor subatomaire fysica, Amsterdam, Netherlands
\item \Idef{org82}Nuclear Physics Group, STFC Daresbury Laboratory, Daresbury, United Kingdom
\item \Idef{org83}Nuclear Physics Institute, Academy of Sciences of the Czech Republic, \v{R}e\v{z} u Prahy, Czech Republic
\item \Idef{org84}Oak Ridge National Laboratory, Oak Ridge, Tennessee, United States
\item \Idef{org85}Petersburg Nuclear Physics Institute, Gatchina, Russia
\item \Idef{org86}Physics Department, Creighton University, Omaha, Nebraska, United States
\item \Idef{org87}Physics Department, Panjab University, Chandigarh, India
\item \Idef{org88}Physics Department, University of Athens, Athens, Greece
\item \Idef{org89}Physics Department, University of Cape Town, Cape Town, South Africa
\item \Idef{org90}Physics Department, University of Jammu, Jammu, India
\item \Idef{org91}Physics Department, University of Rajasthan, Jaipur, India
\item \Idef{org92}Physik Department, Technische Universit\"{a}t M\"{u}nchen, Munich, Germany
\item \Idef{org93}Physikalisches Institut, Ruprecht-Karls-Universit\"{a}t Heidelberg, Heidelberg, Germany
\item \Idef{org94}Politecnico di Torino, Turin, Italy
\item \Idef{org95}Purdue University, West Lafayette, Indiana, United States
\item \Idef{org96}Pusan National University, Pusan, South Korea
\item \Idef{org97}Research Division and ExtreMe Matter Institute EMMI, GSI Helmholtzzentrum f\"ur Schwerionenforschung, Darmstadt, Germany
\item \Idef{org98}Rudjer Bo\v{s}kovi\'{c} Institute, Zagreb, Croatia
\item \Idef{org99}Russian Federal Nuclear Center (VNIIEF), Sarov, Russia
\item \Idef{org100}Russian Research Centre Kurchatov Institute, Moscow, Russia
\item \Idef{org101}Saha Institute of Nuclear Physics, Kolkata, India
\item \Idef{org102}School of Physics and Astronomy, University of Birmingham, Birmingham, United Kingdom
\item \Idef{org103}Secci\'{o}n F\'{\i}sica, Departamento de Ciencias, Pontificia Universidad Cat\'{o}lica del Per\'{u}, Lima, Peru
\item \Idef{org104}Sezione INFN, Bari, Italy
\item \Idef{org105}Sezione INFN, Bologna, Italy
\item \Idef{org106}Sezione INFN, Cagliari, Italy
\item \Idef{org107}Sezione INFN, Catania, Italy
\item \Idef{org108}Sezione INFN, Padova, Italy
\item \Idef{org109}Sezione INFN, Rome, Italy
\item \Idef{org110}Sezione INFN, Trieste, Italy
\item \Idef{org111}Sezione INFN, Turin, Italy
\item \Idef{org112}SSC IHEP of NRC Kurchatov institute, Protvino, Russia
\item \Idef{org113}SUBATECH, Ecole des Mines de Nantes, Universit\'{e} de Nantes, CNRS-IN2P3, Nantes, France
\item \Idef{org114}Suranaree University of Technology, Nakhon Ratchasima, Thailand
\item \Idef{org115}Technical University of Split FESB, Split, Croatia
\item \Idef{org116}The Henryk Niewodniczanski Institute of Nuclear Physics, Polish Academy of Sciences, Cracow, Poland
\item \Idef{org117}The University of Texas at Austin, Physics Department, Austin, Texas, USA
\item \Idef{org118}Universidad Aut\'{o}noma de Sinaloa, Culiac\'{a}n, Mexico
\item \Idef{org119}Universidade de S\~{a}o Paulo (USP), S\~{a}o Paulo, Brazil
\item \Idef{org120}Universidade Estadual de Campinas (UNICAMP), Campinas, Brazil
\item \Idef{org121}University of Houston, Houston, Texas, United States
\item \Idef{org122}University of Jyv\"{a}skyl\"{a}, Jyv\"{a}skyl\"{a}, Finland
\item \Idef{org123}University of Liverpool, Liverpool, United Kingdom
\item \Idef{org124}University of Tennessee, Knoxville, Tennessee, United States
\item \Idef{org125}University of the Witwatersrand, Johannesburg, South Africa
\item \Idef{org126}University of Tokyo, Tokyo, Japan
\item \Idef{org127}University of Tsukuba, Tsukuba, Japan
\item \Idef{org128}University of Zagreb, Zagreb, Croatia
\item \Idef{org129}Universit\'{e} de Lyon, Universit\'{e} Lyon 1, CNRS/IN2P3, IPN-Lyon, Villeurbanne, France
\item \Idef{org130}V.~Fock Institute for Physics, St. Petersburg State University, St. Petersburg, Russia
\item \Idef{org131}Variable Energy Cyclotron Centre, Kolkata, India
\item \Idef{org132}Warsaw University of Technology, Warsaw, Poland
\item \Idef{org133}Wayne State University, Detroit, Michigan, United States
\item \Idef{org134}Wigner Research Centre for Physics, Hungarian Academy of Sciences, Budapest, Hungary
\item \Idef{org135}Yale University, New Haven, Connecticut, United States
\item \Idef{org136}Yonsei University, Seoul, South Korea
\item \Idef{org137}Zentrum f\"{u}r Technologietransfer und Telekommunikation (ZTT), Fachhochschule Worms, Worms, Germany
\end{Authlist}
\endgroup

\end{document}